\documentclass[aps,prd,twocolumn,floatfix,nofootinbib,showpacs,superscriptaddress]{revtex4-1}

\usepackage[mathscr,scaled=1.15]{urwchancal}
\DeclareFontFamily{OT1}{pzc}{}
\DeclareFontShape{OT1}{pzc}{m}{it}%
{<-> s * [1.15] pzcmi7t}{}
\DeclareMathAlphabet{\mathpzc}{OT1}{pzc}{m}{it}

\usepackage{amsmath,amsfonts,amssymb,bm}

\usepackage{longtable}

\usepackage{extarrows}

\usepackage{graphicx}

\usepackage{color}

\usepackage{slashed}

\usepackage{pifont}

\definecolor{purple}{rgb}{0.5,0,0.5}
\definecolor{blue}{rgb}{0.0,0,0.9}

\begin{document}

\title{A Bridge from Euclidean  Nonperturbative approach to Minkowskian Distribution Functions}

\author{Langtian Liu}
\affiliation{Department of Physics and State Key Laboratory of Nuclear Physics and Technology,
Peking University, Beijing 100871, China}
\affiliation{Collaborative Innovation Center of Quantum Matter, Beijing 100871, China}

\author{Lei Chang}
\affiliation{School of Physics, Nankai University, Tianjin 300071, China}

\author{Yu-xin Liu }
\email[Corresponding author: ]{yxliu@pku.edu.cn}
\affiliation{Department of Physics and State Key Laboratory of Nuclear Physics and Technology,
Peking University, Beijing 100871, China}
\affiliation{Collaborative Innovation Center of Quantum Matter, Beijing 100871, China}
\affiliation{Center for High Energy Physics, Peking University, Beijing 100871, China}

\date{\today}

\begin{abstract}
We give out a simple way to connect the parton distribution functions defined in Minkowskian space and the nonperturbative QCD methods grounded in Euclidean space (e.g., lattice QCD(LQCD), Dyson-Schwinger (DS) equations, functional renormalization group (FRG) approach) in this work.
We combine the MIT bag model with the DS equation approach to calculate the  longitudinal distribution function, transverse distribution function and scalar distribution function in a proton at renormalization point $\mu = 2\,\text{GeV}$. We look then insight into the dressed effects on the axial, tensor and scalar charges in a nucleon to some extent. This method can be regard as a new bridge between the Euclidean non-perturbative approaches and the Minkowskian space physics.
\end{abstract}

\maketitle

\section{\label{sec1}Introduction}

The parton distribution functions (PDFs) play an essential role in hadron physics, high energy collider physics, and even in the physics beyond standard model~\cite{Barone:2001sp,Gao:2017yyd,Aoki:2019cca}.
The corresponding charges (axial, tensor, scalar charges) are significant in describing hadron properties and structure, and important in interpreting weak interaction results, even probing new physics~\cite{Aoki:2019cca}.
For the high energy collider physics, one can calculate the distribution functions perturbatively~\cite{RevModPhys.67.157}.
However, in the medium or low energy region, the coupling constant of QCD
runs into nonperturbative region thus the perturbative calculation fails.
One needs then to compute the distribution functions nonperturbatively.
Almost all the nonperturbative calculation methods, such as the lattice QCD (LQCD), the Dyson-Schwinger (DS) equation approach, and the functional renormalization group (FRG) approach, are grounded in the Euclidean space, while the distribution functions are usually defined in the Minkowskian space.
One way to solve the problem partly is implementing the quasi-PDFs~\cite{Belitsky:2005qn,Ji:2013PRL,Xiong:2013bka,Ma:2014jla} which equates the corresponding PDFs as the hadron momentum goes to infinity and can be calculated in LQCD.
In the DSE framework, the PDFs in mesons have been calculated (see, e.g., Refs.~\cite{Ding:2019lwe,Ding:2015rkn,Chang:2014lva,Chang:2014gga,Chen:2016sno,Xu:2018eii}),
but the PDFs in nucleon have not yet been evaluated.
In this paper, we search an alternative and much simpler way to look insight into the dressed effects on the parton distribution functions as well as their corresponding charges in nucleon qualitatively.

In the field of nuclear physics or hadron physics,
the MIT bag model is a quite simple, effective and successive model in describing hadron properties
and illustrates the confinement character directly at the beginning of model construction~\cite{PhysRevD.9.3471,PhysRevD.10.2599,PhysRevD.27.1556,1984_book_nuclear}.
The bag model has been implemented to calculate the parton distribution functions in nucleon~\cite{PhysRevD.11.1953,PhysRevD.36.1344,PhysRevD.40.2832,PhysRevD.44.2653,PhysRevD.87.034009}.
For the functional nonperturbative QCD approaches, the DS equation approach~\cite{Roberts:1994dr,ALKOFER2001281,Roberts:20003,Bashir:2012CTP} can characterize
the characters of the dynamical chiral symmetry breaking (DCSB) of QCD~\cite{Pagels:1979hd,Roberts:1990PRD,Roberts:1996NPA,Roberts:1997PRC,Roberts:2003PRC,QCDPT-DSE11,QCDPT-DSE12,Alkofer:2009AP,QCDPT-DSE13} and the quark confinement from the quark spectral functions~\cite{Bhagwat:2002tx,Alkofer:2003jj,Fischer:2009gk,Qin:2011PRD,Qin:2013ufa,Gao:2016qkh}.
Even though these two approaches can give out some properties and some predictions of hadrons~\cite{1984_book_nuclear,ALKOFER2001281,PhysRevD.12.2060,Roberts2016jp,Qin:2011xq,Chen:2016bpj,Chang:2013PRL1,Chang:2013PRL2,Chen:2015mga,Gao:2017PRD}, both of them have their own weaknesses.
The MIT bag model can show the confinement effect directly but few dressed effects.
On the other hand, the DS equation approach can show the dressed effects
apparently~\cite{Roberts:1990PRD,Bashir:2012CTP,Roberts:1996NPA,Roberts:1997PRC,Roberts:2003PRC,QCDPT-DSE11,QCDPT-DSE12,Alkofer:2009AP,Alkofer:2003jj,Fischer:2009gk,Qin:2011PRD,Qin:2013ufa,Gao:2016qkh,Bhagwat:2002tx,ALKOFER2001281,Roberts2016jp,Qin:2011xq,Chen:2016bpj,Chang:2013PRL1,Chang:2013PRL2,Chen:2015mga,Gao:2017PRD,Fischer:2003rp,Maris:2005tt,Aguilar:2009nf,Eichmann:2011vu,QCDPT-DSE13}.
Therefore it is natural to combine these two approaches to simply look insight into the dressed effects of the parton distribution functions in hadrons.
In the scheme of DS equations, great efforts have been made to develop the sophisticated truncation scheme and incorporate the dressed effects (see, e.g., Refs.~\cite{Munczek:1994zz,Bender:2002PRC,Bhagwat:2004,Chang:2009zb,Fischer:2009AP,CLR:2011,QCLR:2013,ABIP:1418,Williams:2015EPJA,BCPQR:2017,GTL:2017,TGL:2019,Chang:2009zb}),
however the most commonly used truncation scheme is still the rainbow-ladder (RL) approximation which is just the leading order truncation.
Since we have extended the truncation beyond RL approximation by utilizing the Munczek's quark-gluon vertex approximation~\cite{Munczek:1994zz,Liu:2019wzj},
we would like to investigate the dressed effects not only in leading order but also beyond leading order in this paper.

In this work, we combine the MIT bag model with the DS equation approach to calculate the quark distribution functions as well as their corresponding charges in a nucleon, e.g. the longitudinal distribution function, the transversity distribution function and the scalar distribution function.
We compare the distribution functions as well as the proton isovector charges with or without the dressed effects and analyze the dressed effects in RL approximation and beyond the RL approximation at the same pion dacay constant $f_{\pi}^{}$ in chiral limit.
Aiming to investigate more dressed effects on the distribution functions in a nucleon,
we add more and more dressed effects step by step.
At first, we calculate the quark vertexes with only the first two-particle irreducible (2PI) scattering kernel in the Bethe-Salpeter (BS) equation.
Secondly, we restore the full 2PI scattering kernel to calculate the quark vertexes in the BS equation.
All the calculations within the DS equation approach are done at renormalization point $\mu  = 2\text{GeV}$.

We find that the dressed effects vary on different parton distribution functions.
For the longitudinal distribution function, we found the axial coupling $g_{a}^{}$ we calculated under the bare axial vector vertex is close to the experimental value.
The dressed effects in the RL approximation lower the longitudinal distribution and in case of beyond the RL approximation, more dressed effects will increase the longitudinal distribution function at the same pion decay constant $f_{\pi}^{}$.
%
%
For the transversity distribution, the dressed effects in the RL truncation lower the transversity distribution.
When more and more dressed effects are taken into account, the transversity distribution decreases continuously at first but then rises up under the same pion decay constant $f_{\pi}^{}$ condition.
As for the scalar distribution, the dressed effects in the RL truncation rise the scalar distribution function up. When more and more dressed effects are taken into account, the scalar distribution is lowered much comparing to the RL truncation at the same pion decay constant $f_{\pi}^{}$.
%
%
We may infer then that the dressed scalar charge would be fixed to a value.
In general, the dressed effects make the distribution functions in a nucleon more convincing.

The remainder of this article is organized as the follows. In Sec.\ref{sec-dist-bag} we describe how to calculate the distribution functions in the MIT bag model.
In Sec.\ref{sec-combine-bag-dse} we take the dressed effects into account and combine the MIT bag model with the DS equation approach to calculate the distribution functions in a proton.
%
%
We give finally a summary in Sec.\ref{sec-sum}.

\section{Distribution Functions in the Bag Model}
\label{sec-dist-bag}

Let us first describe how to calculate the distribution functions in the bag model \cite{PhysRevD.36.1344,PhysRevD.44.2653,PhysRevD.87.034009,PhysRevD.11.1953,PhysRevD.40.2832} in order to combine the bag model with the DS equation approach.

\subsection{Quark Field in the Bag}

The cornerstone of the bag model is the idea of bag and the wave function of a fermion in the bag is the solution of the free massless Dirac equation in a spherical cavity with radius $R$~\cite{Sakurai:2011zz}:
\begin{equation}
\varphi_\lambda(\bm x, t) = N
\begin{bmatrix}
j_0\left( \frac{\omega_{n,\kappa}r}{R} \right) U_\lambda\\
i \bm \sigma \cdot \hat {\bm x} j_1 \left( \frac{\omega_{n,\kappa}r}{R} \right) U_\lambda
\end{bmatrix}
e^{-i \frac{\omega_{n \kappa} t}{R}}\,,
\end{equation}
where $\lambda$ denotes the spin polarization direction, $r = |\bm x|$ and $\hat x = \bm x / r$.
Here $j_i(x)$ is the $ith$ order spherical Bessel function and we have $ER = \omega_{n \kappa}$,
with $E$ the energy of the fermion and $R$ the bag radius.
$\omega_{n \kappa}$ can be determined with the boundary condition $j_{0}^{}(\omega_{n \kappa}) = j_{1}^{}(\omega_{n \kappa})$. Since we focus on the ground state of hadrons,
in the following we fix it to the first solution of the boundary equation and omit the subscripts $n, \kappa$:
$$\Omega = \omega_{1,-1} = 2.04\;.$$
$U_\lambda$ is the harmonic spin wave function,
\begin{equation}
U_\lambda = \frac{1}{\sqrt{4\pi}}
\begin{bmatrix}
\delta_{\lambda, \uparrow}\\
\delta_{\lambda, \downarrow}
\end{bmatrix}\,,
\end{equation}
and $N$ is the normalization constant,
\begin{equation}
\begin{split}
&\int d^3x\; \varphi_\lambda^\dagger(\bm x, t) \varphi_\lambda(\bm x, t) = 1 \\
&\Rightarrow N^2 = \frac{1}{R^3}\frac{\Omega^4}{\Omega^2 - \sin^2(\Omega)}\;.
\end{split}
\end{equation}
As for a hadron consists of $n$ quarks in the bag model, the hadron energy $E_H$ would be
\begin{equation}
E_H = \sum_{i=1}^n \frac{\Omega_i}{R} + \frac{4 \pi}{3}R^3 B - \frac{Z}{R}\,,
\label{eq-energy}
\end{equation}
where $R$, $B$ is the  bag radius, the bag constant, respectively.
$\Omega_{i}$ is the parameter $\Omega$ for $ith$ quark.
The last term in Eq.\eqref{eq-energy} is the correction on the center of motion and the zero point energies.
The hadron mass are determined from the minimal energy condition
$\frac{\partial E_{H}^{}}{\partial R}\Big|_{E_{H}^{}=M} = 0$.
For a nucleon, $n=3$, we get then the nucleon mass as
\begin{equation}
M = \frac{4 \Omega - 4 Z / 3}{R}\,.
\end{equation}
By fitting the hadron mass spectrum, one can get the value of the parameters.
We adopt the values listed in Table.~\ref{tab-bag-para} which are quoted from Refs.~\cite{1984_book_nuclear,PhysRevD.12.2060}.
In the following calculations, we always make use of these parameters.
\begin{table}[h]
\caption{The bag model parameters (taken from Refs.~\cite{1984_book_nuclear,PhysRevD.12.2060}) we take in
our numerical calculations.}
\label{tab-bag-para}
\begin{ruledtabular}
\begin{tabular}{ccccc}
$B$ & $R$ & $M$ & $\Omega$ & $Z$ \\
\hline
$\left(0.146\,\text{GeV}\right)^4$ & $1.0313\,\textrm{fm}$ & $1.0922\,$GeV & $2.04$ & $1.84$ \\
\end{tabular}
\end{ruledtabular}
\end{table}

Before we calculate the distribution functions, we need the quark wave function in the momentum space. The quark wave function in the momentum space within the bag model is
\begin{equation}
\varphi(\bm k, t) = 4\pi R^3 N
\begin{bmatrix}
t_{00}(y) U_m\\
\bm \sigma \cdot \hat{\bm k} t_{11}(y) U_m
\end{bmatrix}
e^{-i \frac{\Omega t}{R}}\,,
\end{equation}
where $y = k R, k = |\bm k|, \hat {\bm k} = \bm k/k$ and
\begin{equation}
t_{i\, j}(y) = \int_0^1 du u^2 j_i(yu) j_j(\Omega u)\,, \qquad y = kR\,.
\end{equation}
The results of $t_{00}(y), t_{11}(y)$ have been shown in Appendix. \ref{appendix-wave-fourier}

With the quark wave function, a quark field centered at $\bm a$ in the bag can be expanded as
\begin{equation}
\psi(\bm x, t) = \sum_{\lambda = \uparrow, \downarrow} a_{f_\lambda}(\bm a)\varphi_\lambda(\bm x - \bm a, t) + \cdots
\end{equation}
where $a^\dagger_{f_{\lambda}}(\bm a)$ ($a_{f_{\lambda}}(\bm a)$) is the creation (annihilation) operator
of a $f$ flavor quark centered at $\bm a$ with spin $\lambda$, and ``$\cdots$'' represents other quark states that we could neglect in this work.
The creation and annihilation operators satisfy relation:
\begin{equation}
\left\{a_i(\bm a), a^\dagger_j(\bm b)\right\} = \delta_{ij} \int d^3x \varphi^\dagger_j(\bm x - \bm b) \varphi_i(\bm x - \bm a)\, ,
\end{equation}
in which the right hand side is the overlap integral between two quarks in the bag
and the result is showed in Appendix.~\ref{appendix-overlap-int}.

A proton consists of 2 valence $u$ quarks and 1 valence $d$ quark.
According to the standard quark model, the wave function of a spin-up proron can be written as~\cite{Thomson:2013zua}:
\begin{equation}
\begin{split}
|P\uparrow\rangle = \frac{1}{\sqrt{18}}&\left(2 u_\uparrow u_\uparrow d_\downarrow - u_\uparrow u_\downarrow d_\uparrow - u_\downarrow u_\uparrow d_\uparrow \right.\\
&\left.+2u_\uparrow d_\downarrow u_\uparrow - u_\uparrow d_\uparrow u_\downarrow - u_\downarrow d_\uparrow u_\uparrow \right.\\
&\left.+2d_\downarrow u_\uparrow u_\uparrow - d_\uparrow u_\uparrow u_\downarrow - d_\uparrow u_\downarrow u_\uparrow\right)\,.
\end{split}
\end{equation}
Then, in the bag model, the wave  function of a spin-up proton reads
\begin{equation}
\begin{split}
|P\uparrow,&\, \bm r = \bm a\rangle \\=  \frac{1}{\sqrt{18}}
&\left(2 a^\dagger_{u_\uparrow}(\bm a) a^\dagger_{u_\uparrow}(\bm a) a^\dagger_{d_\downarrow}(\bm a) \right.\\
&\left.- a^\dagger_{u_\uparrow}(\bm a) a^\dagger_{u_\downarrow}(\bm a) a^\dagger_{d_\uparrow}(\bm a) \right.\\
&\left.- a^\dagger_{u_\downarrow}(\bm a) a^\dagger_{u_\uparrow}(\bm a) a^\dagger_{d_\uparrow}(\bm a) + \cdots \right) |0,\, \bm r = \bm a\rangle\,,
\end{split}
\end{equation}
where $|0,\, \bm r = \bm a\rangle$ is the empty bag centered at $\bm r = \bm a$
and satisfies $\langle 0,\, \bm{r} = \bm{a}|0,\, \bm{r} = \bm{b} \rangle = \delta_{\bm{a}\bm{b}}$.
This is the proton wave function in coordinate space and we will take the Peierls--Yoccoz (PY) projection method to get the proton wave function in momentum space~\cite{PhysRevD.24.1416,PhysRev.89.1102,Peierls_1957}.

\subsection{PY Projection Method}

For a general static hadron state $\vert H_{B}(\bm x)\rangle$ centered at $\bm x$,
we can decompose it in terms of plane-wave momentum eigenstates $|H(\bm p)\rangle$ \cite{PhysRevD.24.1416}:
\begin{equation}
|H_B(\bm x)\rangle = \int \frac{d^3p}{(2\pi)^3}e^{i \bm p \cdot \bm x}\left[\frac{\phi(\bm p)}{W_H(p)}\right] |H(\bm p)\rangle\,.
\end{equation}
The inverse relation is
\begin{equation}
|H(\bm p)\rangle = \left[\frac{W_H(p)}{\phi(\bm p)}\right]\int d^3x e^{-i \bm x \cdot \bm p}|H_B(\bm x)\rangle\,,
\label{eq-decom}
\end{equation}
where $W_{H}(\bm p)$ is the normalization factor for the plane-wave
\begin{equation}
\langle H(\bm p)|H(\bm p')\rangle = (2\pi)^3\delta^{(3)}(\bm p - \bm p') W_H(p)\,.
\label{eq-norm}
\end{equation}
The physical results do not depend on the choice of $W_{H}$~\cite{PhysRevD.24.1416} and
we will omit this normalization factor in the following.
Substituting Eq.~\eqref{eq-decom} into Eq.~\eqref{eq-norm} we get
\begin{equation}
|\phi(\bm p)|^2 = \int d^3r e^{-i \bm r \cdot \bm p} \langle H_B(\bm 0)|H_B(\bm r)\rangle\,.
\label{eq-component}
\end{equation}
In the case of a proton, we have explicitly
\begin{equation}
|\phi(\bm p)|^2 = |\phi_3(\bm p)|^2\,,
\end{equation}
where
\begin{equation}
|\phi_n(\bm p)|^2 = \int d^3a  e^{-i \bm p \cdot \bm a} \left[ \int d^3x \varphi^\dagger (\bm x - \bm a) \varphi(\bm x) \right]^n\,.
\label{eq-overlap}
\end{equation}
The calculation of this integral is showed in Appendix.~\ref{appendix-overlap-int}.

\subsection{Quark Distribution Function in the Bag}

The quark distribution function of a nucleon in the bag model is defined as~\cite{PhysRevD.87.034009}:
\begin{equation}
\begin{split}
q_i(x) &= 2M \int \frac{d \xi^+}{4\pi}e^{-i \frac{M x}{\sqrt{2}} \xi^+}\times\\
&\langle N;\bm p = 0| \bar \psi_i(\xi) \gamma^+ \psi_i(0)|N; \bm p =0 \rangle\big|_{\xi^-, \xi_\perp = 0}\,,
\end{split}
\label{eq-quark-dist}
\end{equation}
where $M$ is the mass of the nucleon and $\xi^{\pm} = \frac{\xi^0 \pm \xi^3}{\sqrt{2}}$,
$\gamma^{+} = \frac{\gamma^0 + \gamma^3}{\sqrt{2}}$.
We shall note that $-1\leq x \leq 1$ is the momentum fraction.
Let us now denote
\begin{equation}
\mathcal{M} = \langle N;\bm p = 0| \bar \psi_i(\xi) \gamma^+ \psi_i(0)|N; \bm p =0 \rangle\,,
\end{equation}
we can rewrite $\mathcal{M}$ with Eq.~\eqref{eq-decom} as
\begin{equation}
\begin{split}
\mathcal{M} = \frac{1}{|\phi_3(0)|^2}&\int d^3a d^3b \times \\
& \langle N; \bm r = \bm a|\bar \psi_i(\xi) \gamma^+ \psi_i(0)|N; \bm r = \bm b\rangle\,.
\end{split}
\end{equation}
We denote the probability that one picks up a quark with flavor $f$ and spin polarization $\lambda$
from a nucleon as $\langle N;\bm r = 0| P_{f,\lambda}|N; \bm r = 0\rangle$.
For example, in a longitudinally polarized proton, the probabilities of finding a quark with flavor $f$
and spin parallel or anti-parallel to the longitudinal polarization direction are listed in Table~\ref{tab-prob} (see Ref.~\cite{PhysRevD.87.034009} for details).
\begin{table}[h]
\caption{The probabilities of finding a quark with flavor $f$ and polarization direction $\lambda$ in the longitudinal polarized proton.}
\label{tab-prob}
\begin{tabular}{c|c|c}
\hline \hline
~~~~~$f$~~~~~ & ~~~~~$\lambda$~~~~~ & ~~~~~$P_{f,\lambda}$~~~~~\\
\hline
$u$ & $\uparrow$ & 5/3\\
\hline
$u$ & $\downarrow$ & 1/3\\
\hline
$d$ & $\uparrow$ & 1/3\\
\hline
d & $\downarrow$ & 2/3 \\
\hline \hline
\end{tabular}
\end{table}
As for the transversely polarized proton, the probability of find a quark with flavor $f$ and spin polarized direction parallel or anti-parallel to the proton transversely polarized direction are the same within the valence quark language.

One can then get
\begin{equation}
\begin{split}
& \langle N;\bm a| \bar \psi_i(\xi) \gamma^+ \psi_i(0) | N; \bm b\rangle
= \sum_\lambda \langle N; \bm 0| P_{f,\lambda}|N; \bm 0\rangle \times \\
& \qquad |\phi_2(\bm b - \bm a)|^2 \bar \varphi(\bm \xi - \bm  a) \gamma^+ \varphi(-\bm b) \, e^{i \frac{\Omega \xi^0}{R}}\,.
\end{split}
\label{eq-matele}
\end{equation}
One can further express the functions in  momentum space as
\begin{subequations}
\begin{align}
& |\phi_2(\bm b - \bm a)|^2 = \int \frac{d^3k_1}{(2\pi)^3}e^{i \bm k_1 \cdot (\bm b - \bm a)}|\phi_2(\bm k_1)|^2\,,\\
& \bar \varphi(\bm \xi - \bm a) = \int \frac{d^3k_2}{(2\pi)^3} e^{-i \bm k_2 \cdot (\bm \xi - \bm a)} \bar \varphi(\bm k_2)\,, \\
& \varphi(- \bm b) = \int \frac{d^3k_3}{(2\pi)^3}e^{i \bm k_3 \cdot (- \bm b)} \varphi(\bm k_3)\, .
\end{align}
\end{subequations}
And one can get the result of $\mathcal{M}$ as:
\begin{equation}
\begin{split}
\mathcal{M} &= \sum_\lambda \langle N; \bm 0| P_{f,\lambda}|N; \bm 0\rangle \int \frac{d^3k}{(2\pi)^3} e^{i (\Omega \xi^0/R - \bm k \cdot \bm \xi)} \times\\
&\qquad\quad \bar\varphi(\bm k) \gamma^+ \varphi(\bm k) \frac{|\phi_2(\bm k)|^2}{|\phi_3(\bm 0)|^2}
\,.
\end{split}
\end{equation}
Finally, we arrive at the quark distribution function of a nucleon in the bag model as
\begin{widetext}
\begin{equation}
\begin{split}
q_i(x) &= \left(\sum_\lambda \langle N; \bm 0| P_{f,\lambda}|N; \bm 0\rangle\right)\frac{M}{2\pi}\int d \xi^+ e^{i (-\frac{M x}{\sqrt{2}} + \frac{\Omega/R - k_3}{\sqrt{2}})\xi^+} \int \frac{d^3 k}{(2\pi)^3} \bar \varphi(\bm k) \gamma^+ \varphi(\bm k) \frac{\phi_2(|\bm k)|^2}{|\phi_3(\bm 0)|^2}\,, \\
&= \left(\sum_\lambda \langle N; \bm 0| P_{f,\lambda}|N; \bm 0\rangle\right)M \int \frac{k dk d \varphi d k_3}{(2\pi)^3} \delta(-\frac{Mx}{\sqrt{2}} + \frac{\Omega/R - k_3}{\sqrt{2}})\bar\varphi(\bm k) \gamma^+ \varphi(\bm k) \frac{\phi_2(|\bm k)|^2}{|\phi_3(\bm 0)|^2}\,,\\
&= \sqrt{2}M \left(\sum_\lambda \langle N; \bm 0| P_{f,\lambda}|N; \bm 0\rangle\right) \int_{k_{min}}^\infty \frac{k dk}{(2\pi)^2} \bar\varphi(\bm k) \gamma^+ \varphi(\bm k) \frac{|\phi_2(\bm k)|^2}{|\phi_3(\bm 0)|^2}\,,
\end{split}
\label{eq-quark-dist-res}
\end{equation}
\end{widetext}
where $k_{min} = |\Omega/R - M x|$ and the momentum in longitudinal direction
\begin{equation}
k_{3} = \Omega/R - M x\,,
\end{equation}
from the delta function in second step.
Since
\begin{equation}
\begin{split}
&\bar \varphi(\bm k) \gamma^+ \varphi(\bm k) = \frac{4\pi R^3 \Omega^4}{\sqrt{2}(\Omega^2 - \sin^2 \Omega)} \times\\
&\qquad\left[t_{00}^2(k) + t_{11}^2(k) + 2 \hat k_3 t_{00}(k) t_{11}(k)\right]\,,
\end{split}
\label{eq-inner-prod}
\end{equation}
where $\hat k_3 = k_3/k$,
substituting Eq.~\eqref{eq-inner-prod} into Eq.~\eqref{eq-quark-dist-res} we obtain
\begin{equation}
\begin{split}
&q_i(x) = \frac{4\pi M R^3 \Omega^4}{(\Omega^2 - \sin^2 \Omega)}\left(\sum_\lambda \langle N; \bm 0| P_{f,\lambda}|N; \bm 0\rangle\right)\times\\
& \int_{k_{min}}^\infty \frac{k dk}{(2\pi)^2} \left[t_{00}^2(k) + t_{11}^2(k) + 2 \hat k_3 t_{00}(k) t_{11}(k)\right] \frac{|\phi_2(\bm k)|^2}{|\phi_3(\bm 0)|^2}\,.
\end{split}
\end{equation}

\begin{figure}[]
\centering
\includegraphics[width=0.40\textwidth]{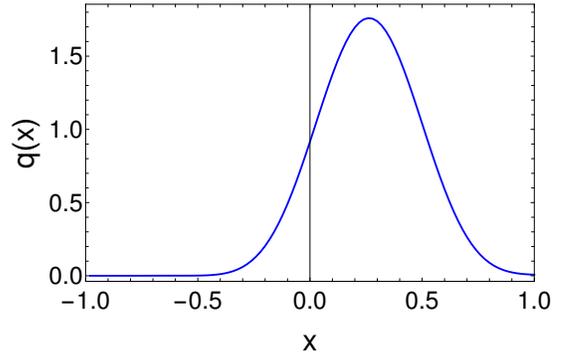}
\caption{The single quark distribution function $q(x)$ of a nucleon in the bag model.}
\label{fig-quark-dist}
\end{figure}

In Fig.~\ref{fig-quark-dist} we desplay the single quark distribution function in a nucleon within the bag model. For the numerical calculation, the parameters which we take are those listed in Table.~\ref{tab-bag-para}.

By integrating the single quark distribution function we get the valence quark number
%
$$ \int_{-1}^1 dx \; q(x) \approx 1 \, . $$

\section{Combining Bag Model with the DS Equation Approach}
\label{sec-combine-bag-dse}

In this section, we combine the MIT bag model with the Dyson-Schwinger equation approach to get a bridge that can connect the physics between the Euclidean space and the Minkowskian space.
Before we calculate the dressed distribution functions, let us describe the key points of the DS equation approach of QCD we take in this work firstly.

\subsection{Truncation Scheme of the DS Equation Approach}
\label{approx-scheme}

It is known that the DS equations are a set of integral equations composed of infinite number but countable coupled quark, gluon, ghost and corresponding interaction vertex equations.
For now, the truncation scheme about the DS equations mainly talking about the approximation of the quark-gluon vertex. Different truncation schemes lead to different BS equations.
Let's denote a dressed quark vertex as $\Gamma(k; 0)$ regardless of the Lorentz indices.
The general form of the BS equation with the dressed quark vertex can be written as
\begin{equation}
\label{eq-bs}
\begin{split}
[\Gamma(k;0)]_{EF} &= Z_v [\Gamma_0(k;0)]_{EF} \\
&\quad + \int_{dq}^\Lambda [K(k,q;0)]^{GH}_{EF}[\chi(q;0)]_{GH}\, ,
\end{split}
\end{equation}
where $Z_v$ is the renormalization coefficient for the dressed quark vertex and inhomogeneous term $\Gamma_0(k;0)$ is the tree level term of the dressed quark vertex. $K(k, q; 0)$ is the 2PI scattering kernel and depends on the truncation scheme. $\chi(q;0)=S(q)\Gamma(q;0)S(q)$ denotes the unamputated dressed quark vertex where $ S(q) $ is the quark propagator with momentum $ q $ and $E,F,G,H$ are spinor indices here.

For RL approximation, the quark gluon vertex takes the form
\begin{equation}
\Gamma_\nu^a(p+k,p) =  t^a \gamma_\nu\,,
\end{equation}
and the 2PI scattering kernel is
\begin{equation}
[K(k,q;0)]^{GH}_{EF} = -Z_1 g^2D_{\mu\nu}(q-k)t^a\left[\gamma_\mu\right]_{EG}t^a\left[\gamma_\nu\right]_{HF}\, ,
\end{equation}
where $ t^a $ is the color SU(3) group generator and $ Z_1 $ is the quark-gluon vertex renormalization coefficient. Here $ D_{\mu\nu}(k) $ is the gluon propagator with momentum $ k $.
In the DS equation approach, the gluon propagator in Landau gauge takes the form
\begin{equation}
g^{2} D_{\alpha \beta}(k) = \mathcal{G}(k^2)(\delta_{\alpha \beta} - \frac{k_{\alpha} k_{\beta}}{k^2})\, ,
\end{equation}
where
\begin{align}
&\mathcal{G}(k^2) =\mathcal{G}_{\text{ir}}^{}(k^2) + \mathcal{G}_{\text{uv}}^{}(k^2) \, , \\
&\mathcal{G}_{\text{ir}}(k^2) = \frac{8 \pi^2}{\omega^5} m_{g}^{3} e^{-k^2/\omega^2} \, , \\
&\mathcal{G}_{\text{uv}}(k^2) = \frac{8 \pi^2 \gamma_m}{\ln[\tau+(1+k^2/\Lambda_{QCD}^2)^2]} \frac{1-e^{-k^2/4m_t^2}}{k^2}\, ,
\end{align}
$\omega$ and $m_{g}$ are the interaction parameters of the infrared gluon propagator,
denoting the interaction width and the magnitude, respectively.
In the ultraviolet gluon propagator,
$\gamma_{m} = 12/(33-2n_f)$ is the anomalous dimension and $n_{f}$ is the number of flavors.
In our calculations, we take $n_{f} = 4$.
$\Lambda_{\text{QCD}}$ is the QCD scale that characterizes the non-perturbative QCD and we choose $\Lambda_{\text{QCD}} = 0.234\,\text{GeV}$.
$\tau = e^{2}$ with $e$ being the natural constant, and $m_{t} = 0.5\,\text{GeV}$.

For the framework coupling the BS equation and the DS equation (DSE-BSE) based on the Munczek's quark-gluon vertex approximation~\cite{Liu:2019wzj}, the Munczek's quark-gluon vertex approximation takes as
\begin{equation}
\label{vertex}
i \Gamma_\nu^{a}(p+k,p) =t^{a} \frac{\partial}{\partial p^\nu} \int_0^1 d \alpha S^{-1}(p+\alpha k) + i \Gamma_\nu^{aT}(p+k,p)\, ,
\end{equation}
where $\Gamma_\nu^{aT}(p+k,p)$ is the transverse part of the quark-gluon vertex which satisfies $k^\nu \Gamma_\nu^{aT}(p+k,p) = 0$.
The Munczek's quark-gluon vertex satisfies the Ward-Takahashi identity.
The full scattering kernel under the Munczek's quark-gluon vertex model can be written as
\begin{equation}
[K(k,q;0)]^{GH}_{EF} =  \text{\ding{172}} + \text{\ding{173}}\, ,
\end{equation}
with
\begin{subequations}
	\begin{align}
	&\text{\ding{172}}=-Z_1g^2D_{\mu \nu}(q-k)t^a\left[\gamma_\mu\right]_{EG}t^a\left[\Gamma_\nu(q,k)\right]_{HF}\, ,\\
	\nonumber&\text{\ding{173}}=Z_1g^2\int_{dl}^\Lambda D_{\mu \nu}(l-k) t^a \left[  \gamma_\mu\right] _{EM}\left[  S(l)\right] _{MN}t^a\\
	&\quad\quad\times \frac{\partial}{i\partial k^\nu}\int_0^1d \alpha \left[  S^{-1}(k+\alpha(l-k))\right] _{NG}\\
	\nonumber&\quad\quad\times \delta^{(4)}\left(  k+\alpha(l-k)-q\right) \left[  S^{-1}\left(  k+\alpha(l-k)\right) \right] _{HF}\,.
	\end{align}
\end{subequations}
During computing the dressed quark vertexes within the Munczek's model,
we require that the pion decay constant should be the same as that under the RL approximation.
The pion decay constant in the chiral limit can be calculated from the solution of quark DS equation in the Euclidean space~\cite{Pagels:1979hd}, which reads
\begin{equation}
\begin{split}
f_{\pi}^{2} = & \frac{3}{4\pi^2} \int_0^\infty d p^2 \frac{p^2 B(p^2)}{A^2(p^2) \left[p^2+\frac{B^2(p^2)}{A^2(p^2)}\right]} \times \\
&  \left[\frac{B(p^2)}{A(p^2)} - \frac{p^2}{2} \frac{B'(p^2) A(p^2) - B(p^2) A(p^2)}{A^2(p^2)}\right]\,.
\end{split}
\end{equation}
The gluon parameters in calculating the dressed quark vertexes (axial vector, tensor and scalar quark vertexes) are listed in Table~\ref{tab-gluon-para}.
\begin{table}[h]
	\caption{The gluon parameters we take in calculating the dressed quark vertexes within the DSE-BSE approach.}
	\label{tab-gluon-para}
	\begin{ruledtabular}
		\begin{tabular}{cccc}
			Truncation scheme & $m_{g}/\text{GeV}$ & $\omega/\text{GeV}$ &  $f_\pi/\text{GeV}$\\
			\hline
			RL &  0.82 & 0.5  & 0.0856\\
			\hline
			Munczek  & 0.5766 & 0.5 & 0.0856
		\end{tabular}
	\end{ruledtabular}
\end{table}
Also, in this work, we neglect the second term of scattering kernel in the Munzcek's model at first step and we denote it as \textbf{Mun. $1_{\text{st}}$ truncation}.
Then we restore the full scattering kernel and denote it as \textbf{Mun. full truncation}. By this way, we address the dressed effects step by step.

Having described the truncation schemes in the DSE-BSE approach,
now let us turn to calculate the dressed distribution functions.

\subsection{Longitudinal Distribution Function}

The longitudinal distribution $g(x)$ describes the net quark density of  longitudinally polarized quark in a longitudinally polarized hadron, i.e., the number density of the quark with momentum fraction $x$ and spin parallel to the hadron longitudinal polarization direction minus the density of the quark with the same momentum fraction but spin antiparallel~\cite{Barone:2001sp}.  The axial charge can be read from the integration of the longitudinal distribution $g(x)$  via~\cite{PhysRevLett.67.552,JAFFE1992527}:
\begin{equation}
\triangle q = \int_{-1}^1 dx g(x) = \int_0^1 dx \big[g(x) + \bar g(x)\big]\, ,
\end{equation}
where $\bar g(x)$  denotes the longitudinal distribution of an antiquark in the nucleon.
The relation $g(-x) = \bar g(x), \; x > 0$ is employed in the second step.

\subsubsection{The Longitudinal Distribution Function in a Nucleon under Bare Vertex}

\begin{figure}[h]
\centering
\includegraphics[width=0.43\textwidth]{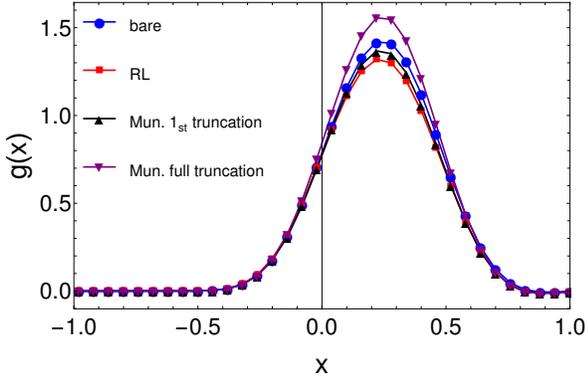}
\caption{Calculated single quark longitudinal distribution function $g(x)$ in a longitudinally polarized proton. {\it Blue line with filled circles} is the result under the bare axial vector vertex.
{\it Red line with filled squares} stands for that under the dressed axial vector vertex in RL approximation. {\it Black line with filled up-triangles} denotes that via the dressed axial vector vertex of only the first scattering kernel in the Munczek's quark gluon vertex model.
And the {\it purple line with filled down-triangles} corresponds to that under the full scattering kernel of Munczek's model.}
\label{fig-axial-dist-dress}
\end{figure}

Suppose a proton is longitudinally polarized along the $z$ direction, then the longitudinal distribution function of a single quark under bare axial vertex in the proton within the bag model is written as
\begin{equation}
\begin{split}
&g_{q}^{}(x) = 2M \int \frac{d \xi^+}{4\pi}e^{-i M x \xi^+}\times\\
&\langle N;\bm p = 0; S| \bar \psi_i(\xi)  \gamma^{+} \gamma_5 \psi_i(0)|N; \bm p =0; S \rangle\big|_{\xi^-, \xi_\perp = 0}\,.
\end{split}
\end{equation}
Just similar to the procedure of calculating the quark distribution function $q(x)$, the result of bare axial vector vertex is
\begin{equation}\label{l-dist-tree}
\begin{split}
g_q^{}(x) &= \sqrt{2}M \left(\sum_\lambda \langle N; \bm 0; S| P_{f,\lambda}|N; \bm 0; S\rangle\right)\times\\
&\quad \int_{k_{min}}^\infty \frac{k dk}{(2\pi)^2} \bar\varphi(\bm k) \gamma^+ \gamma_5 \varphi(\bm k) \frac{|\phi_2(\bm k)|^2}{|\phi_3(\bm 0)|^2}\,.
\end{split}
\end{equation}
Since
\begin{equation}
\begin{split}
&\bar \varphi(\bm k)  \gamma^{+} \gamma_5 \varphi(\bm k) = \frac{4\pi R^3 \Omega^4}{\sqrt{2}(\Omega^2 - \sin^2 \Omega)}\times\\
&\left[t_{00}^2(k) + (2 \hat k_3^2 - 1) t_{11}^2(k) + 2 \hat k_3 t_{00}(k) t_{11}(k)\right]\,,
\end{split}
\end{equation}
we can get the longitudinal distribution function of a single quark in the case of bare axial vector vertex as:
\begin{equation}
\begin{split}
&g_{q}(x) = \frac{4\pi M R^3 \Omega^4}{(\Omega^2 - \sin^2 \Omega)}\left( \langle N; \bm 0; S| P_{f,\lambda}|N; \bm 0; S\rangle\right)\times\\
& \qquad\qquad\int_{k_{min}}^\infty \frac{k dk}{(2\pi)^2} \left[t_{00}^2(k) + (2 \hat k_3^2 - 1) t_{11}^2(k) \right.\\
&\left.\qquad\qquad\qquad+ 2 \hat k_3 t_{00}(k) t_{11}(k)\right] \frac{|\phi_2(\bm k)|^2}{|\phi_3(\bm 0)|^2}\,.
\end{split}
\end{equation}
The obtained result of the single quark longitudinal distribution function is illustrated in Fig.~\ref{fig-axial-dist-dress} (the blue line with filled circle).
Integrating out the longitudinal distribution function we get the single quark axial charge in a nucleon under the bare axial vector vertex as:
\begin{equation}
\triangle q = \int_{-1}^1 dx \, g(x) = 0.7758\,.
\end{equation}
The axial coupling constant is then $g_{A}^{} = 5/3*\triangle q = 1.293$,
which is quite close to the experimental value: $g_{A}^{\text{Exp}} = 1.2727$~\cite{Tanabashi:2018oca}.
%
%

\subsubsection{Dressed Effects on the Longitudinal Distribution Function}
\label{l-dist-sec}

We now look insight into the dressed effects on the longitudinal distribution function.

In principle, we need to consider the dressed effects of both the quarks and the vertexes in the proton
within MIT bag model.
As one knows, in the procedure of calculating the quark distribution function (Eq.~\eqref{eq-quark-dist-res}), the quark wave function comes from the solution of free Dirac equation and because we can not get the full analytic solution of the Dirac equation with interactions,
as a compromise, we consider only the dressed effects of the vertexes which appear in the definition of the distribution functions.

\begin{figure}[htb]
\includegraphics[width=0.43\textwidth]{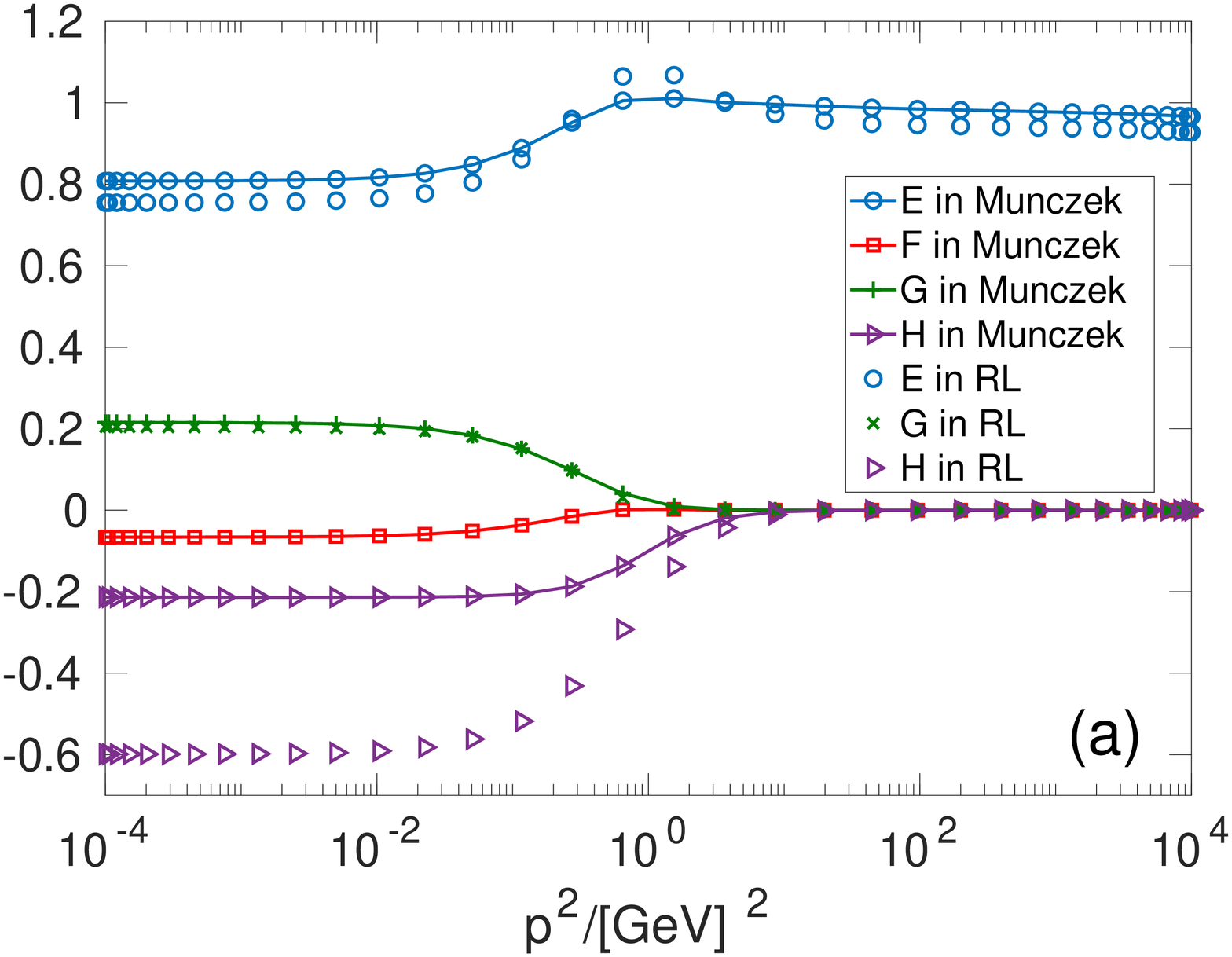}
%
%
\includegraphics[width=0.43\textwidth]{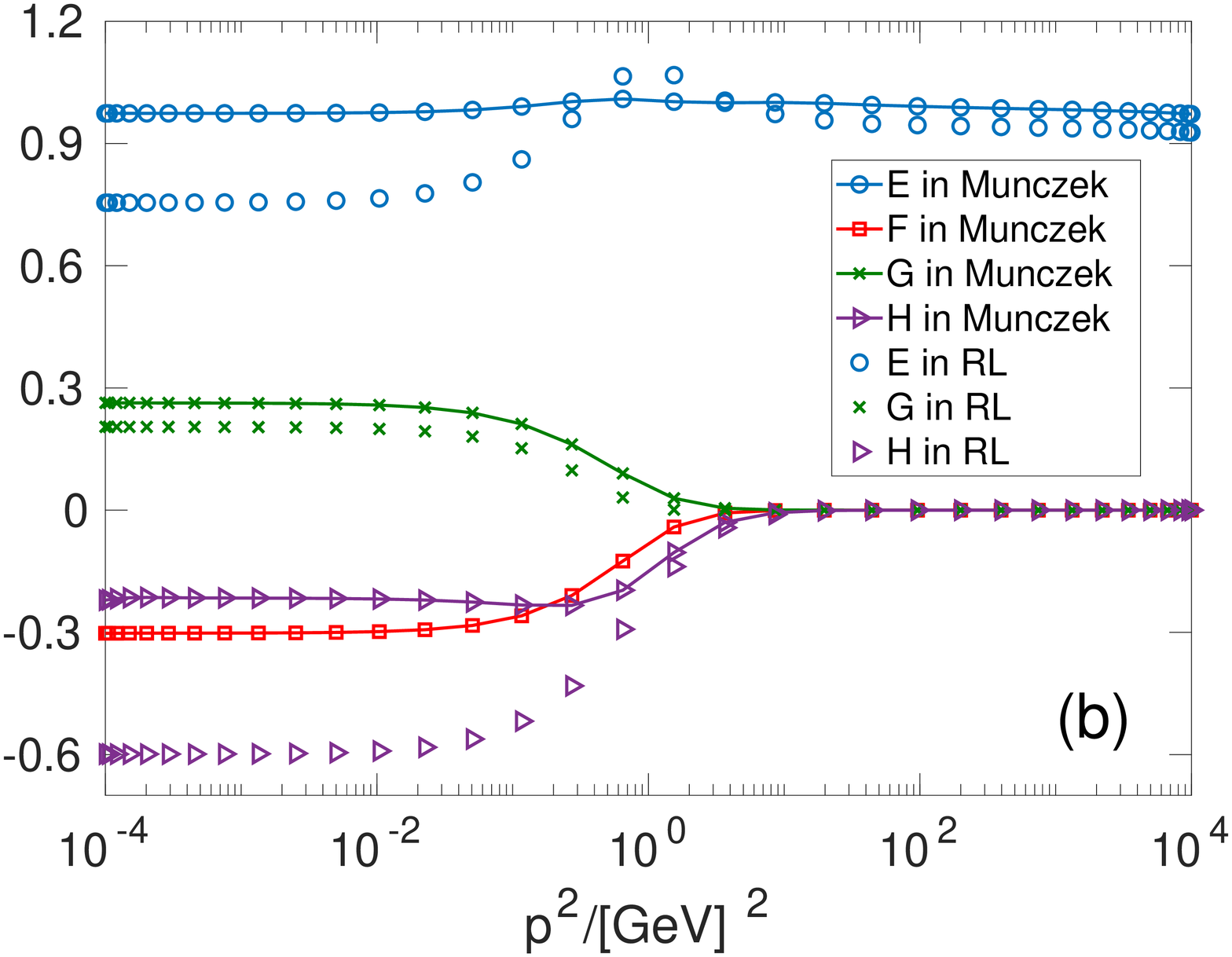}
\caption{Calculated quark axial vertex under the RL approximation and the Munczek's quark gluon vertex approximation. (a) The first scattering kernel in the Munczek's approximation.
(b) The full scattering kernel in the Munczek's approximation.
}
\label{fig-av-sol}
\end{figure}

One way to get the effects of the dressed vertexes is the DS equation approach.
Here we calculate the dressed quark axial vector vertex in the RL approximation
and the Munczek's quark-gluon vertex model~\cite{Munczek:1994zz, Liu:2019wzj}.
The corresponding full axial vector vertex of the longitudinal distribution function is $\Gamma^{\mu}_5$, whose tree level form is $\gamma^{\mu}\gamma_5$. The full quark axial vector vertex can be decomposed as
\begin{equation}
\begin{split}
\Gamma^{\mu}_5(k; 0) & = \gamma^{\mu} \gamma_5 E(k^2) + \big{[}\gamma^{\mu} \gamma_{5}, i\slashed k \big{]}  F(k^{2})\\
& + \left\{ \gamma^{\mu} \gamma_{5}, i\slashed k \right\} G(k^2) +  i\slashed k \gamma^{\mu} \gamma_{5}  i\slashed k H(k^{2})\,,
 \end{split}
 \label{eq-paraz}
\end{equation}
in the Euclidean momentum space. The solutions of the dressed axial vector vertex under the RL approximation and the Munczek's quark-gluon vertex approximation are depicted in Fig.~\ref{fig-av-sol}.
During our calculation, we have taken the renormalization scale as $\mu = 2\, \text{GeV}$.

In view of the dressed effects of the quark axial vector vertex,
the result of dressed longitudinal distribution function in the bag model should be rewritten from Eq.\eqref{l-dist-tree} as
\begin{equation}\label{l-dist-dressed}
\begin{split}
g_q^{}(x) &= \sqrt{2}M \left(\sum_\lambda \langle N; \bm 0; S| P_{f,\lambda}|N; \bm 0; S\rangle\right)\times\\
&\quad \int_{k_{min}}^\infty \frac{k dk}{(2\pi)^2} \bar\varphi(\bm k) \Gamma^{+}_5(k; 0) \varphi(\bm k) \frac{|\phi_2(\bm k)|^2}{|\phi_3(\bm 0)|^2}\,,
\end{split}
\end{equation}
where
\begin{equation}
\begin{split}
\Gamma^{+}_5(k; 0) & = \gamma^{+} \gamma_5 E(k^2) + \big{[}\gamma^{+} \gamma_{5}, i\slashed k \big{]}  F(k^{2})\\
& + \left\{ \gamma^{+} \gamma_{5}, i\slashed k \right\} G(k^2) +  i\slashed k \gamma^{+} \gamma_{5}  i\slashed k H(k^{2})\,.
\end{split}
\label{eq-axvparaz}
\end{equation}

Everything seems to be good to proceed, but there is one thing that draws us back.
That is one usually calculates the dressed quark axial vector vertex within the DSE-BSE framework
in Euclidean space.
However our quark wave function in the bag model is obtained in Minkowskian space,
and the distribution functions in a nucleon is defined in Minkowskian space.
Thus we should find a way to connect the physics in these two spaces.

Fortunately, we can circumvent this difficulty in the MIT bag model. We know that when one variable is independent of time, the result should be the same no matter you calculate it in the Minkowskian space or Euclidean space. This is exactly the case in the MIT bag model. The quark wave function in the bag model is a static quark state.
Turning back to the calculation of the quark distribution function in  Eq.~\eqref{eq-quark-dist-res}, we find that, during the calculation procedures, we have integrated out the time variable and so that the final result is independent of time. Thus when calculating the distribution functions in a nucleon with MIT bag model, it does not matter that one chooses to carry out the calculation in Minkowskian space or in Euclidean space.
We can then combine them together --- we derive the formula analytically in the Minkowskian space
and implement the dressed scalar functions from the DSE-BSE framework which are calculated in the Euclidean space.

Next we give out the result for the other terms in Eq.~\eqref{eq-paraz}.
We must note that $\left(i \slashed{k}\right)^E$ in Euclidean space is the same as $\left(\slashed k\right)^M$ in Minkowskian space. As a consequence, we have
\begin{equation}
\bar \varphi(\bm k) \left[\gamma^{+} \gamma_5, \slashed k\right] \varphi(\bm k) = 0 \,,
\end{equation}
\begin{equation}
\begin{split}
&\bar \varphi(\bm k) \left\{\gamma^{+} \gamma_5, \slashed k\right\} \varphi(\bm k) = \frac{4\pi R^3 \Omega^4}{\sqrt{2}(\Omega^2 - \sin^2 \Omega)} \times  \\
& \qquad\left[2(k_0+k_3) t_{00}^2(k) + 2\big{(} k_0(1-2\hat{k}_3^2) - k_{3} \big{)}  \right. \\
&\qquad\;\left. \times t_{11}^2(k) - 4k \big{(} 1 - \hat k_{3}^{2} \big{)} t_{00}(k) t_{11}(k)\right]\,,
\end{split}
\end{equation}
\begin{equation}
\begin{split}
&\bar \varphi(\bm k) \left(\slashed k  \gamma^{+} \gamma_{5} \slashed k\right) \varphi(\bm k)
=  \frac{4\pi R^3 \Omega^4}{\sqrt{2}(\Omega^2 - \sin^2 \Omega)} \Big{[} \big{(} (k_{0} + k_{3} )^{2} \\
&\qquad + k_{3}^{2} - k^{2} \big{)} t_{00}^{2}(k) + \big{(} 4(k_{0} + k_{3} ) k_{3} (1-\hat{k}_{3}^{2}) \\
&\qquad\; +(1 - 2\hat k_{3}^{2})(k^{2} - k_{3}^{2} -(k_{0} + k_{3})^{2}) \big{)} t_{11}^{2}(k) \\
&\qquad\; - 2\hat k_{3} \big{(}(k_{0} + k_{3})^{2} + k^{2} - k_{3}^{2} \big{)} t_{00}(k) t_{11}(k) \Big{]}\,.
\end{split}
\end{equation}
Here we have made use of the fact that the integration of $\hat k_{1}^{2}$ and $\hat k_{2}^{2}$ should be the same as shown in the following Eq.~\eqref{angle-int}. We can see that the term with $\left[\gamma^{+} \gamma_{5}, \slashed k\right]$ does not contribution to the dressed longitudinal distribution function.

When integrating the momentum,
\begin{equation*}
\begin{split}
&d^3k = k^2 \sin \theta dk d \theta d \phi = k dk d k_3 d \phi, \\
 &\hat k_3 = \cos \theta,\; \hat k_1 = \sin \theta \cos \phi,\; \hat k_2 = \sin \theta \sin \phi\,.
 \end{split}
\end{equation*}
We have
\begin{equation}
\begin{split}
\label{angle-int}
\int_0^{2\pi} d \phi \hat k_1^2 &= \int_0^{2\pi} d \phi \hat k_2^2 =\pi  \sin^2 \theta \\
&=  \frac{1}{2} \int_0^{2\pi} d \phi (1 - \hat k_3^2)\,.
\end{split}
\end{equation}
Noting that $k_{0} - k_{3} = Mx,\;k_{0} = \Omega/R$, we get finally the single quark dressed longitudinal distribution in a longitudinally polarized nucleon in the MIT bag model as
\begin{widetext}
\begin{equation}
\begin{split}
g_q(x) = \;&\frac{4\pi M R^3 \Omega^4}{(\Omega^2 - \sin^2 \Omega)}\left( \langle N; \bm 0| P_{f,\lambda}|N; \bm 0\rangle\right)\int_{k_{min}}^\infty \frac{k dk}{(2\pi)^2} \frac{|\phi_2(\bm k)|^2}{|\phi_3(\bm 0)|^2}\times\\
&\left\{\left[t_{00}^2(k) + (2 \hat k_3^2 - 1) t_{11}^2(k) + 2 \hat k_3 t_{00}(k) t_{11}(k)\right] * E(k^2) \right.\\
&\;\left.+\left[2(k_0+k_3) t_{00}^2(k) + 2\left(k_0(1-2\hat{k}_3^2) - k_3\right) t_{11}^2(k)  - 4k\left(1-\hat k_3^2\right) t_{00}(k) t_{11}(k)\right] * G(k^2)\right.\\
&\;\left.+\left[\left((k_0+k_3)^2+k_3^2-k^2\right) t_{00}^2(k) + \left(4(k_0+k_3)k_3(1-\hat{k}_3^2)+(1-2\hat k_3^2)(k^2-k_3^2-(k_0+k_3)^2)\right) t_{11}^2(k) \right.\right.\\
& \qquad\left.\left.- 2\hat k_3 \left((k_0+k_3)^2+k^2-k_3^2\right) t_{00}(k) t_{11}(k)\right] * H(k^2)\right\}\,,
\end{split}
\end{equation}
\end{widetext}
where $\hat k_{3} = (\Omega/R - Mx)/k$.

We plot the calculated dressed longitudinal distribution functions under the RL approximation
and the Munczek's quark-gluon vertex model in Fig.~\ref{fig-axial-dist-dress}.
By integrating the longitudinal distribution functions, we get the axial vector charges as well as the axial coupling $g_{A}^{}  = \frac{5}{3}\triangle q$ in these cases.
The obtained results are listed in Table.~\ref{tab-avchar}.
The errors come from that we vary the gluon parameter $\omega \in [0.49, 0.51]$ with which the pion decay constant is almost the same.
Also, we have listed the experimental data of the axial charge from PDG~\cite{Tanabashi:2018oca} for the convenience of comparison.
The Table indicates apparently that the result we get from bare axial vector vertex approximation is quite close to the experimental data.
\begin{table}[h]
\caption{Calculated axial vector charges in a proton under three different cases: the bare axial vector vertex approximation, the dressed axial vector vertex under the RL approximation and the dressed axial vector vertex under the Munczek's quark-gluon vertex.
We list also the experimental data of the axial charge from PDG~\cite{Tanabashi:2018oca}
for the convenience of comparison.}
\label{tab-avchar}
\begin{ruledtabular}
\begin{tabular}{c|c|c}
~~Approximation~~ & $\triangle q$ & $g_{A}^{} $\\
\hline
bare & 0.7758 & 1.2930\\
\hline
RL & ~~~~$0.7277^{+(69)}_{-(73)}$~~~~ & ~~~~$1.2128^{+(115)}_{-(122)}$~~~~\\
\hline
Mun. $1_{\text{st}}$  & $0.7431^{+(36)}_{-(68)}$  & $1.2358^{+(60)}_{-(113)}$ \\
\hline
Mun. full & $0.8282^{+(17)}_{-(21)}$ & $1.380^{+(28)}_{-(35)}$\\
\hline
Exp. & - & 1.2723(23) \\
\end{tabular}
\end{ruledtabular}
\end{table}

Anyway,
when considering the dressed effect on the longitudinal distribution,
the distribution is lowered a bit first in the RL approximation than that under the bare axial vector vertex.
Then the longitudinal distribution function begins to rise in the case beyond the RL approximation
than that in the RL approximation as we can see the variation from RL approximation to the Mun. $1_{\text{st}}$ truncation (consider only the first scattering kernel) and to the Mun. full truncation (consider the full scattering kernels).
That is, the axial charge decreases at first in the RL approximation and then increases gradually
when taking more dressed effects into account comparing to the bare axial vertex case.

\subsection{Transversity Distribution Function}

The transversity distribution $h(x)$ describes the net quark density of  transversely polarized quark in the transversely polarized hadrons, i.e., the number density of quark with momentum fraction $x$ and spin parallel to the hadron transverse polarization direction minus the density of the quark with the same momentum fraction but spin antiparallel.
%
%
The tensor charge can be read from the integral over the transversity distribution function $h(x)$  via~\cite{PhysRevD.52.2960,PhysRevLett.67.552,JAFFE1992527,Cortes:1991ja}:
\begin{equation}
\delta q = \int_{-1}^1 dx h(x) = \int_0^1 dx \big[h(x) - \bar h(x)\big]\, ,
\end{equation}
where $\bar h(x)$  denotes to the transversity distribution of an antiquark in the nucleon.
The relation $h(-x) = -\bar h(x), \; x > 0$ is employed in the second step.

\subsubsection{Transversity Distribution in a Nucleon under Bare Vertex}

\begin{figure}[h]
\centering
\includegraphics[width=0.45\textwidth]{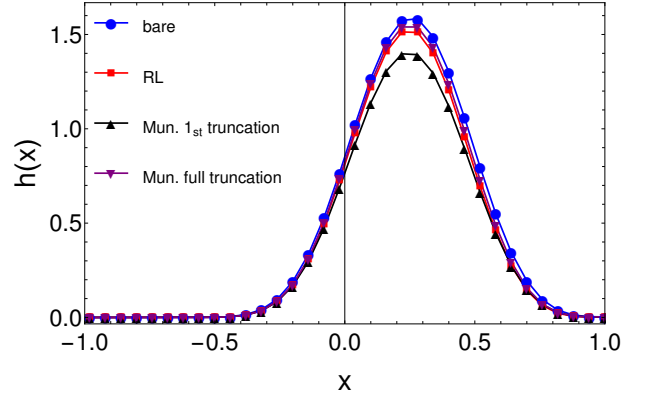}
\caption{Calculated transversity distribution functions in the bare tensor vertex approximation ({\it blue line with filled circles}), in the dressed tensor vector vertex under RL approximation ({\it red line with filled squares}) and the dressed tensor vector vertex under the Munczek's quark gluon vertex model ({\it black line with filled up-triangles for the first scattering kernel case and purple line with filled down-triangles for the full scattering kernel case}).}
\label{fig-ten-dist-dress}
\end{figure}

Suppose a proton is transversely polarized along the $x$ direction, then the transversity distribution function of a single quark in the proton within the bag model is defined as
\begin{equation}
\begin{split}
&h_q(x) = 2M \int \frac{d \xi^+}{4\pi}e^{-i M x \xi^+}\times\\
&\langle N;\bm p = 0; S| \bar \psi_i(\xi) i \sigma^{1+} \gamma_5 \psi_i(0)|N; \bm p =0; S \rangle\big|_{\xi^-, \xi_\perp = 0}\,.
\end{split}
\end{equation}
Here $-1 \leq x \leq 1$ is the momentum fraction.
%
%
Similarly, the result of bare tensor vertex is
\begin{equation}\label{t-dist-tree}
\begin{split}
h_q^{}(x) &= \sqrt{2}M \left(\sum_\lambda \langle N; \bm 0; S| P_{f,\lambda}|N; \bm 0; S\rangle\right)\times\\
&\quad \int_{k_{min}}^\infty \frac{k dk}{(2\pi)^2} \bar\varphi(\bm k) i \sigma^{1+} \gamma_5 \varphi(\bm k) \frac{|\phi_2(\bm k)|^2}{|\phi_3(\bm 0)|^2}\,.
\end{split}
\end{equation}
Since
\begin{equation}
\begin{split}
&\bar \varphi(\bm k) i \sigma^{1+} \gamma_5 \varphi(\bm k)  \\
&= \frac{4\pi R^3 \omega^4}{\sqrt{2}(\omega^2 - \sin^2 \omega)}\left[t_{00}^2(k) + \hat k_3^2 t_{11}^2(k) + 2 \hat k_3 t_{00}(k) t_{11}(k)\right]\,,
\end{split}
\end{equation}
we can get the transversity distribution function of a quark in the case of bare tensor vertex in bag model:
\begin{equation}
\begin{split}
h_q(x) &= \frac{4\pi M R^3 \Omega^4}{(\Omega^2 - \sin^2 \Omega)}\left( \langle N; \bm 0; S| P_{f,m}|N; \bm 0; S\rangle\right)\times\\
&\qquad \int_{k_{min}}^\infty \frac{k dk}{(2\pi)^2} \left[t_{00}^2(k) + \hat k_3^2 t_{11}^2(k) + \right.\\
&\qquad\qquad\left. 2 \hat k_3 t_{00}(k) t_{11}(k)\right] \frac{|\phi_2(\bm k)|^2}{|\phi_3(\bm 0)|^2}\,.
\end{split}
\end{equation}
The calculated result of the single quark transversity distribution function is depicted in Fig.~\ref{fig-ten-dist-dress}.
Integrating out the transversity distribution function we get the single quark tensor charge in the case of bare tensor vertex as:
\begin{equation}
\delta q = \int_{-1}^1 dx \, h(x) = 0.8873\,.
\end{equation}
We should notice that the tensor charge is scale dependent.
The scale dependence can be obtained via perturbative QCD theory as~\cite{Barone:2001sp,Barone:1997fh}:
\begin{equation}
\delta q (Q^{2}) = \delta q(Q_{0}^{2}) \Big[\frac{\alpha_{s}(Q_{0}^{2})}{\alpha_{s}(Q^{2})}\Big]^{-4/(33-2 n_{f})}\, ,
\end{equation}
where $\alpha_{s}(k^{2})$ is the running coupling constant of strong interaction and $n_{f}$ is the number of flavor. As mentioned before, we take $n_{f} = 4$, and we choose here the scale $Q = 2 \,\text{GeV}$.

\subsubsection{Dressed Effects on Transversity Distribution Function}

\begin{figure}[htb]
\includegraphics[width=0.43\textwidth]{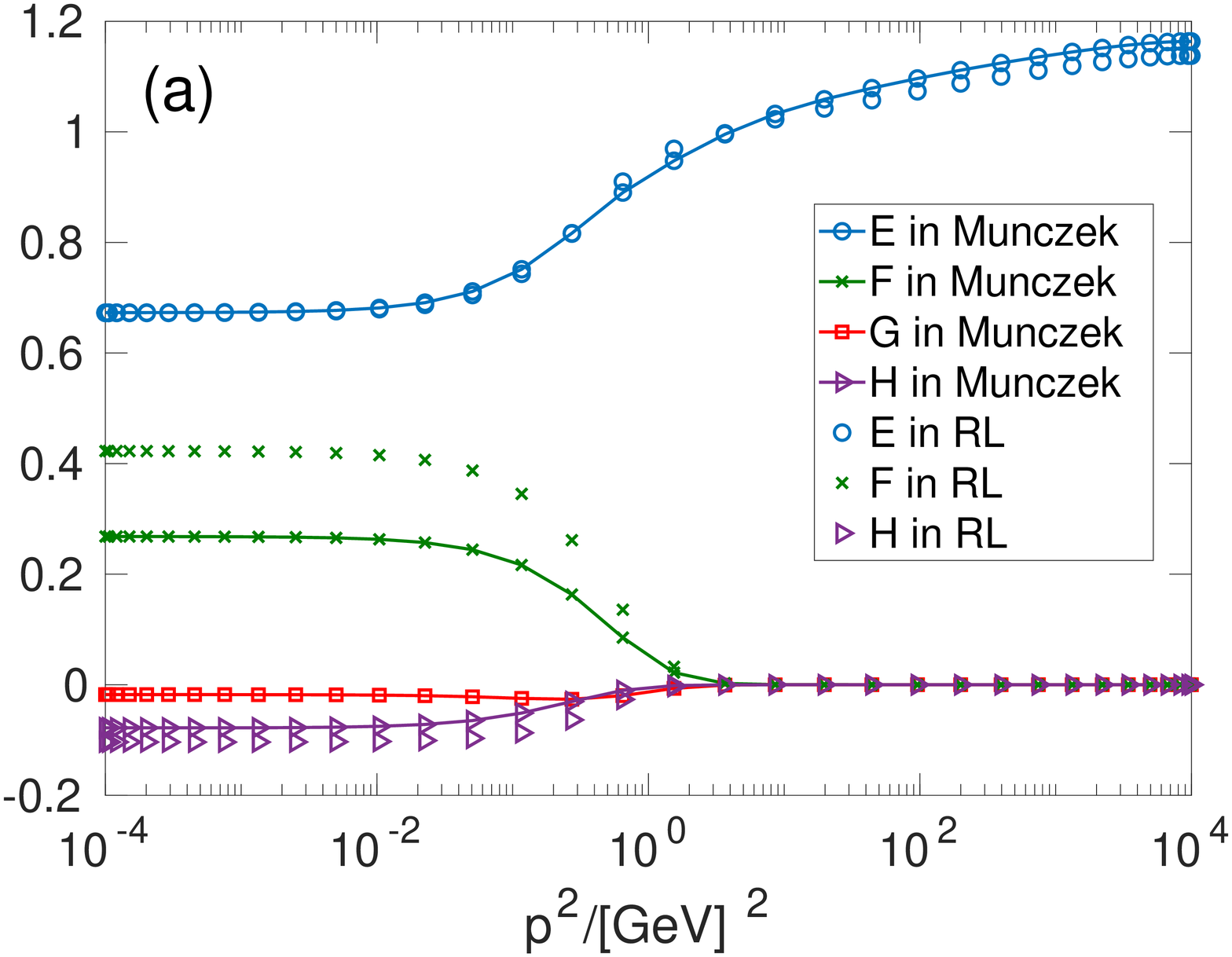}
%
%
\includegraphics[width=0.43\textwidth]{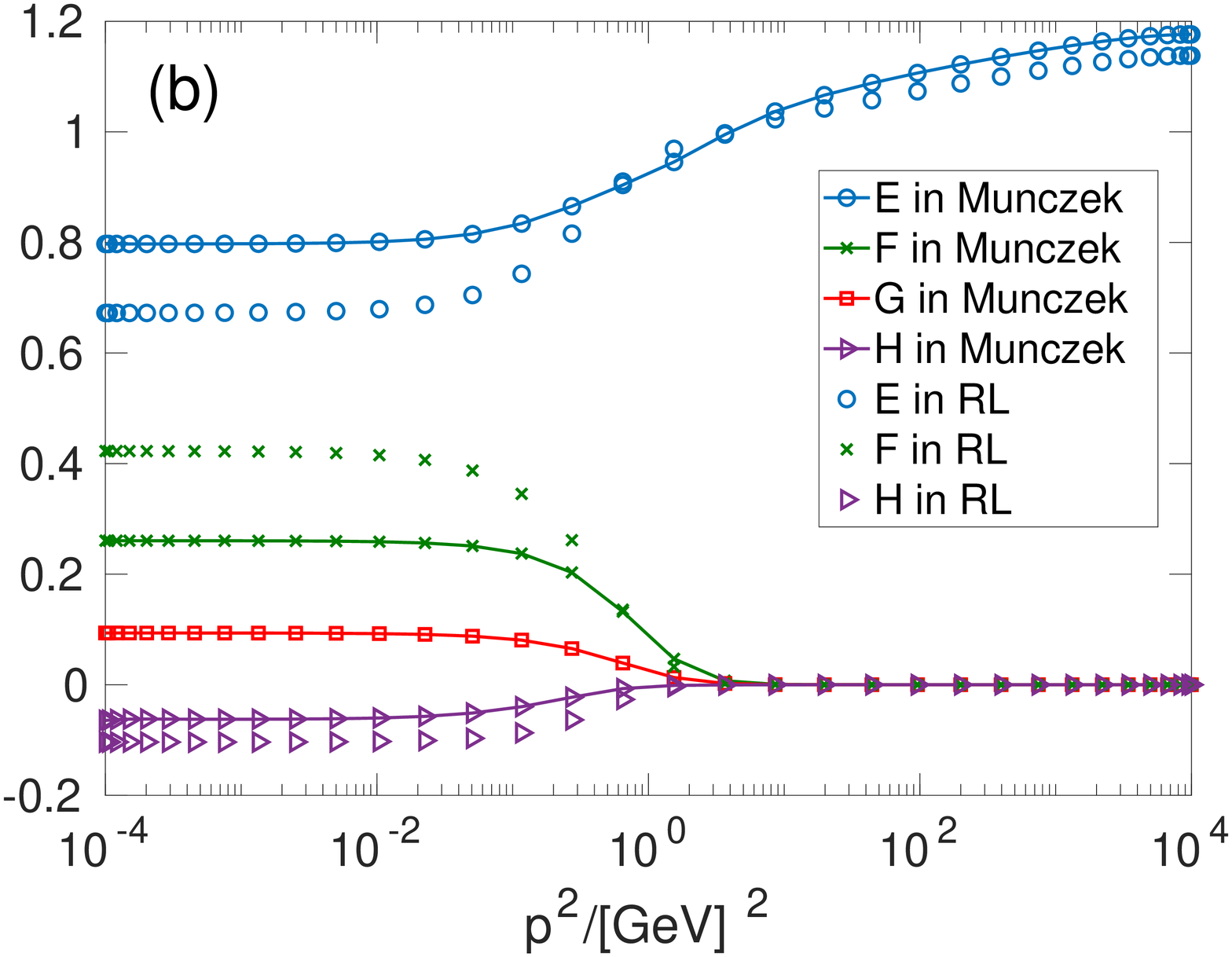}
\caption{Calculated quark tensor vertex under the RL approximation and the Munczek's quark-gluon vertex model. (a)\ with only first scattering kernel in Munczek's model.
(b)\ with the full scattering kernel in Munczek's model. }
\label{fig-tv-sol}
\end{figure}

Similar to the case of longitudinal distribution function, the corresponding full tensor vertex of the transversity distribution function is $i\Gamma^{\mu \nu}_{5}$, whose tree level form
is $i \sigma^{\mu \nu}\gamma_{5}$. We decompose the full quark tensor vertex as
\begin{equation}
\begin{split}
&\Gamma^{\mu \nu}_5(k; 0) = \sigma^{\mu \nu} \gamma_5 * E(k^2) + \left\{ i\slashed k,\sigma^{\mu \nu} \gamma_5\right\} * F(k^2)\\
&\qquad + \left[ i\slashed k, \sigma^{\mu \nu} \gamma_5\right] * G(k^2) +  i\slashed k \sigma^{\mu \nu} \gamma_5  i\slashed k * H(k^2)\,,
 \end{split}
 \label{eq-paraz-tv}
\end{equation}
in the Euclidean momentum space.
We calculate the dressed quark tensor vertex with the same gluon parameters listed in Table~\ref{tab-gluon-para} and show the obtained results in Fig.~\ref{fig-tv-sol}.
During the calculation,
we firstly omit the  second term of scattering kernel in the Munczek's quark gluon vertex model
and then restore it to the full kernel.

In light of the dressed effects of quark tensor  vertex,
the result of dressed transversity distribution function in bag model should be rewritten from Eq.\eqref{t-dist-tree} as
\begin{equation}\label{t-dist-dressed}
\begin{split}
g_q^{}(x) &= \sqrt{2}M \left(\sum_\lambda \langle N; \bm 0; S| P_{f,\lambda}|N; \bm 0; S\rangle\right)\times\\
&\quad \int_{k_{min}}^\infty \frac{k dk}{(2\pi)^2} \bar\varphi(\bm k) i \Gamma^{1+}_5(k; 0) \varphi(\bm k) \frac{|\phi_2(\bm k)|^2}{|\phi_3(\bm 0)|^2}\,,
\end{split}
\end{equation}
where
\begin{equation}
\begin{split}
&\Gamma^{1+}_5(k; 0) = \sigma^{1+} \gamma_5 * E(k^2) + \left\{ i\slashed k,\sigma^{1+} \gamma_5\right\} * F(k^2)\\
&\qquad + \left[ i\slashed k, \sigma^{1+} \gamma_5\right] * G(k^2) +  i\slashed k \sigma^{1+} \gamma_5  i\slashed k * H(k^2)\,.
\end{split}
\label{eq-tvparaz}
\end{equation}
As the same reason as that in calculating the dressed longitudinal distribution function,
the calculation of the transversity distribution function in the bag model is independent of time.
We then derive the formula analytically in the Minkowskian space and make use of the dressed scalar functions from the DS equation calculation results which are calculated in the Euclidean space.

Next we give out the result for the other terms in Eq.~\eqref{eq-paraz-tv}.
\begin{subequations}
\begin{align}
&\label{eq-55a}\bar \varphi(\bm k) \left[\slashed k,  i \sigma^{1+} \gamma_5\right] \varphi(\bm k) = 0 \,,\\
&\label{eq-49b} \bar \varphi(\bm k) \left\{\slashed k,  i \sigma^{1+} \gamma_{5} \right\} \varphi(\bm k) = \frac{4\pi R^3 \Omega^4}{\sqrt{2}(\Omega^2 - \sin^2 \Omega)} \times \nonumber \\
& \qquad\left[2(k_0+k_3) t_{00}^2(k) - 2\left(k_0\hat{k}_3^2 + k_3\right) t_{11}(k)\right. \\
&\nonumber \qquad\;\left. - 2k(1 - \hat k_3^2) t_{00}(k) t_{11}(k)\right] \Big{]} \,,\\
&\bar \varphi(\bm k) \left(\slashed k  i \sigma^{1+} \gamma_5 \slashed k\right) \varphi(\bm k)  =  \frac{4\pi R^3 \Omega^4}{\sqrt{2}(\Omega^2 - \sin^2 \Omega)}\times  \nonumber \\
& \qquad \left[(k_0+k_3)^2 t_{00}^2(k) + \left((k_0+k_3)^2\hat{k}_3^2 \right. \right.\\
&\nonumber\qquad\;\left.\left.+k^2(-\hat k_1^2+\hat k_2^2)^2+4k_3(k_0+k_3)\hat k_1^2\right) t_{11}^2(k) \right.\\
&\nonumber\qquad\; \left.- 2\left(2k_2 \hat k_2 (k_0+k_3)+(k_0+k_3)^2\hat k_3\right) t_{00}(k) t_{11}(k)\right]\,.
\end{align}
\end{subequations}
In the Eq.~\eqref{eq-55a} we have taken the fact that the angle integration of $k_{2}$ turns out to be $0$ since $\chi_m^\dagger \sigma_2 \chi_m = 0$.
We can see that the base $\left[\slashed k,  i \sigma^{1+} \gamma_{5}\right]$ does not contribute to the transversity distribution.

When integrating the momentum we make use of the relation in Eq.~\eqref{angle-int} and
%
$$ \displaylines{\hspace*{1cm}
\hspace*{1cm} \int_0^{2\pi} d \phi \hat k_1^4 = \int_0^{2\pi} d \phi \hat k_2^4 = \frac{3\pi}{4}  \sin^4 \theta \hfill{(64a)} \cr
\hspace*{2.58cm} =  \frac{3}{8} \int_0^{2\pi} d \phi (1 - \hat k_3^2)^2\, , \hfill{} \cr } $$

$$\displaylines{\hspace*{0.4cm}
\int_0^{2\pi} d \phi \hat k_1^2 \hat k_2^2 = \frac{\pi}{4} \sin^4 \theta  = \frac{1}{8}\int_0^{2\pi} d \phi (1-\hat k_3^2)^2\,.  \hfill{(64b)} } $$
Noting that $k_{0} - k_{3} = Mx,\;k_{0} = \Omega/R$, we get finally the dressed transversity distribution
in a nucleon within the MIT bag model as

\setcounter{equation}{64}
\begin{widetext}
\begin{equation}
\begin{split}
h_q(x) = \;&\frac{4\pi M R^3 \Omega^4}{(\Omega^2 - \sin^2 \Omega)}\left( \langle N; \bm 0| P_{f,m}|N; \bm 0\rangle\right)\int_{k_{min}}^\infty \frac{k dk}{(2\pi)^2} \frac{|\phi_2(\bm k)|^2}{|\phi_3(\bm 0)|^2}\times\\
&\left\{\left[t_{00}^2(k) + \hat k_3^2 t_{11}^2(k) + 2 \hat k_3 t_{00}(k) t_{11}(k)\right] * E(k^2) \right.\\
&\;\left.+\left[2(k_0+k_3) t_{00}^2(k) - 2\left(k_0\hat{k}_3^2 + k_3\right) t_{11}(k) - 2k (1-\hat k_3^2) t_{00}(k) t_{11}(k)\right] * F(k^2)\right.\\
&\;\left.+\left[(k_0+k_3)^2 t_{00}^2(k) + \left((k_0+k_3)^2\hat{k}_3^2+k^2(1-\hat k_3^2)^2/2 +2k_3(k_0+k_3)(1-\hat k_3^2)\right) t_{11}^2(k)\right.\right.\\
&\qquad\left.\left. - 2\left(k (1-\hat k_3^2) (k_0+k_3) +(k_0+k_3)^2\hat k_3\right) t_{00}(k) t_{11}(k)\right] * H(k^2)\right\}\,,
\end{split}
\end{equation}
\end{widetext}
where $\hat k_{3} = (\Omega/R - Mx)/k$.

\begin{table*}[]
\caption{Calculated tensor charge in a proton under four different cases: the bare tensor vertex,
the dressed tensor vertex under the RL approximation and the dressed tensor vertex in Munczek quark-gluon vertex model.}
\label{tab-tenchar}
%
\centering
	\begin{ruledtabular}
		\begin{tabular}{ccccc}
			Approximation & $\delta q$ & $\delta u$ & $\delta d$ & $g_{T}^{} $ \\
			\hline
			bare & $0.8873$ & $1.183$ & $-0.2958$ & $1.479$ \\
			\hline
			RL & $0.8203^{+(91)}_{-(89)}$ & $1.094^{+(12)}_{-(12)}$ & $-0.2734^{+(30)}_{-(30)}$ & $1.367^{+(15)}_{-(15)}$\\
			\hline
			Mun.$1_{\text{st}}$ & $0.7724^{+(24)}_{-(25)}$ & $1.030^{+(32)}_{-(33)}$ & $-0.2575^{+(8)}_{-(8)}$ & $1.287^{+(40)}_{-(42)}$\\
			\hline
			Mun. full & $0.8437^{+(111)}_{-(107)}$  & $1.125^{+(148)}_{-(143)}$  &
			$-0.2812^{+(37)}_{-(36)}$ & $1.406^{+(185)}_{-(178)}$ \\
			\hline
			LQCD & $-$ & $0.790(27)$ & $-0.198(10)$ & $0.989(32)$ \\
			\hline
			Faddeev& $-$ & $0.912^{+(42)}_{-(47)}$ & $-0.218^{+(4)}_{-(5)}$ & $1.130^{+(42)}_{-(47)}$\\
			\hline
			Exp. & $-$ & $0.39(10)$ & $-0.11(26)$ & $0.53(25)$ \\
		\end{tabular}
	\end{ruledtabular}
\end{table*}

We illustrate the obtained dressed transversity distribution functions under the RL approximation and the Munczek quark-gluon vertex model in Fig.~\ref{fig-ten-dist-dress}.
By integrating the transversity distribution functions, we get the tensor charge in the four cases and show the obtained results in Table~\ref{tab-tenchar}.
The errors come from that we vary the gluon parameter $\omega \in [0.49, 0.51]$ with which the pion decay constant maintains almost the same.
As a comparison, we list also the experimental data of the tensor charges~\cite{Anselmino:2013vqa,Ye:2016prn,Radici:2018iag}, the results obtained via LQCD calculation~\cite{Yamanaka:2018uud,Gupta:2018qil,Bhattacharya:2016zcn,Alexandrou:2017qyt}
and Faddeev equation calculation~\cite{Wang:2018kto}.

We can notice from Fig.~\ref{fig-ten-dist-dress} and Table~\ref{tab-tenchar} that,
when we consider the dressed effects on the transversity distribution,
the distribution is lowered in the RL approximation and continues lowering in Mun. $1_{\text{st}}$ truncation (consider only first scattering kernel).
However the transversity distribution function gets risen up in the Mun. full truncation
(restore the full two scattering kernels).
Accordingly, the tensor charge decreases gradually when taking the dressed effects into account
in the RL and the Mun. $1_{\text{st}}$ truncation, but it rises up in the Mun. full truncation.

\subsection{Scalar Distribution Function}

Analogous to the axial vector charge and the tensor charge, we define the scalar charge as the integral of the scalar distribution function, which corresponds to the scalar quark vertex.
With the scalar distribution function being denoted as $s(x)$, the quark scalar charge can be obtained by the integration
\begin{equation}
s_q = \int_{-1}^1 dx s(x)\,.
\end{equation}
Since, in the isospin symmetry limit (see Ref.~\cite{Aoki:2019cca} page 237),
\begin{equation}
\langle p\big| \bar u \Gamma d \big| n\rangle = \langle p\big| \bar u \Gamma u - \bar d \Gamma d \big | p\rangle\,,
\end{equation}
thus the single quark scalar charge is also the isovector scalar coupling in the bag model, which reads
\begin{equation}
g_{s}^{} = s_{u} - s_{d} = 2 s_{q} - s_{q} = s_{q}  \, .
\end{equation}
Here $\big{|} p \rangle$, $\big{|} n\rangle$ is the proton, the neutron state, respectively.
$u$, $d$ is the $u$ quark, $d$ quark, respectively. $\Gamma$ is the quark vertex.

\subsubsection{Scalar Distribution in Nucleon under Bare Vertex}

\begin{figure}[]
\centering
\includegraphics[width=0.43\textwidth]{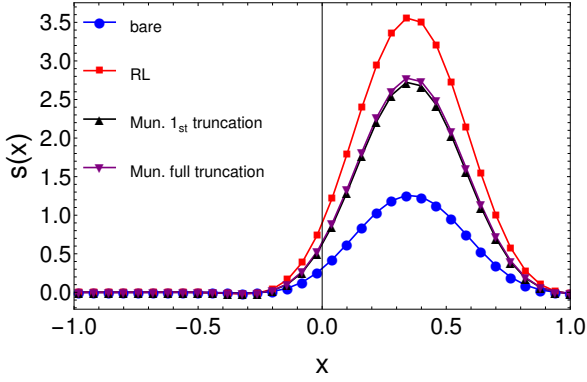}
\caption{Calculated scalar distribution function in the cases of bare scalar vertex ({\it blue line with filled circles}), dressed scalar vertex under RL approximation ({\it red line with filled squares})
and the dressed scalar vertex under the Munczek's quark-gluon vertex model ({\it black line with filled up-triangles for the first scattering kernel case and purple line with filled down-triangles for the full scattering kernel case}).}
\label{fig-scalar-dist-dress}
\end{figure}

In the case of bare scalar vertex, the single quark scalar distribution function in the bag model is defined as
\begin{equation}
\begin{split}
&s(x) = 2M \int \frac{d \xi^+}{4\pi}e^{-i M x \xi^+}\times\\
&\langle N;\bm p = 0; S| \bar \psi_i(\xi) \cdot 1 \cdot \psi_i(0)|N; \bm p =0; S \rangle\big|_{\xi^-, \xi_\perp = 0}\, ,
\end{split}
\end{equation}
where $-1 \leq x \leq 1$ is the momentum fraction. Actually, the scalar distribution function is independent of nucleon polarization and we can then choose any polarization direction in derivation.
Here we simply assume the proton is transversely polarized along $x$ direction. The result of scalar distribution function under bare scalar quark vertex in bag model is
\begin{equation}\label{s-dist-tree}
\begin{split}
s_q^{}(x) &= \sqrt{2}M \left(\sum_\lambda \langle N; \bm 0; S| P_{f,\lambda}|N; \bm 0; S\rangle\right)\times\\
&\quad \int_{k_{min}}^\infty \frac{k dk}{(2\pi)^2} \bar\varphi(\bm k) \cdot 1 \cdot \varphi(\bm k) \frac{|\phi_2(\bm k)|^2}{|\phi_3(\bm 0)|^2}\,.
\end{split}
\end{equation}
Since
\begin{equation}
\begin{split}
&\bar \varphi(\bm k)\cdot 1 \cdot \varphi(\bm k) = \frac{4\pi R^3 \Omega^4}{\sqrt{2}(\Omega^2 - \sin^2 \Omega)}\times\\
&\qquad\left[t_{00}^2(k) - t_{11}^2(k)\right]\,,
\end{split}
\end{equation}
we obtain the single quark scalar distribution function $s(x)$ in the bag model under bare scalar vertex as:
\begin{equation}
\begin{split}
&s(x) = \frac{4\pi M R^3 \Omega^4}{(\Omega^2 - \sin^2 \Omega)}\left( \langle N; \bm 0; S| P_{f,m}|N; \bm 0; S\rangle\right)\times\\
& \int_{k_{min}}^\infty \frac{k dk}{(2\pi)^2} \left[t_{00}^2(k) - t_{11}^2(k) \right] \frac{|\phi_2(\bm k)|^2}{|\phi_3(\bm 0)|^2}\,.
\end{split}
\end{equation}
The calculated result of the single quark scalar distribution function is displayed in Fig.~\ref{fig-scalar-dist-dress} (blue line with filled circles). Integrating out the scalar distribution function we get the single quark scalar charge in the case of bare scalar vertex as:
\begin{equation}
s_{q} = \int_{-1}^1 dx \, s(x) = 0.6655\,.
\end{equation}

\subsubsection{Dressed Effects on the Scalar Distribution Function}

\begin{figure}[htb]
\includegraphics[width=0.43\textwidth]{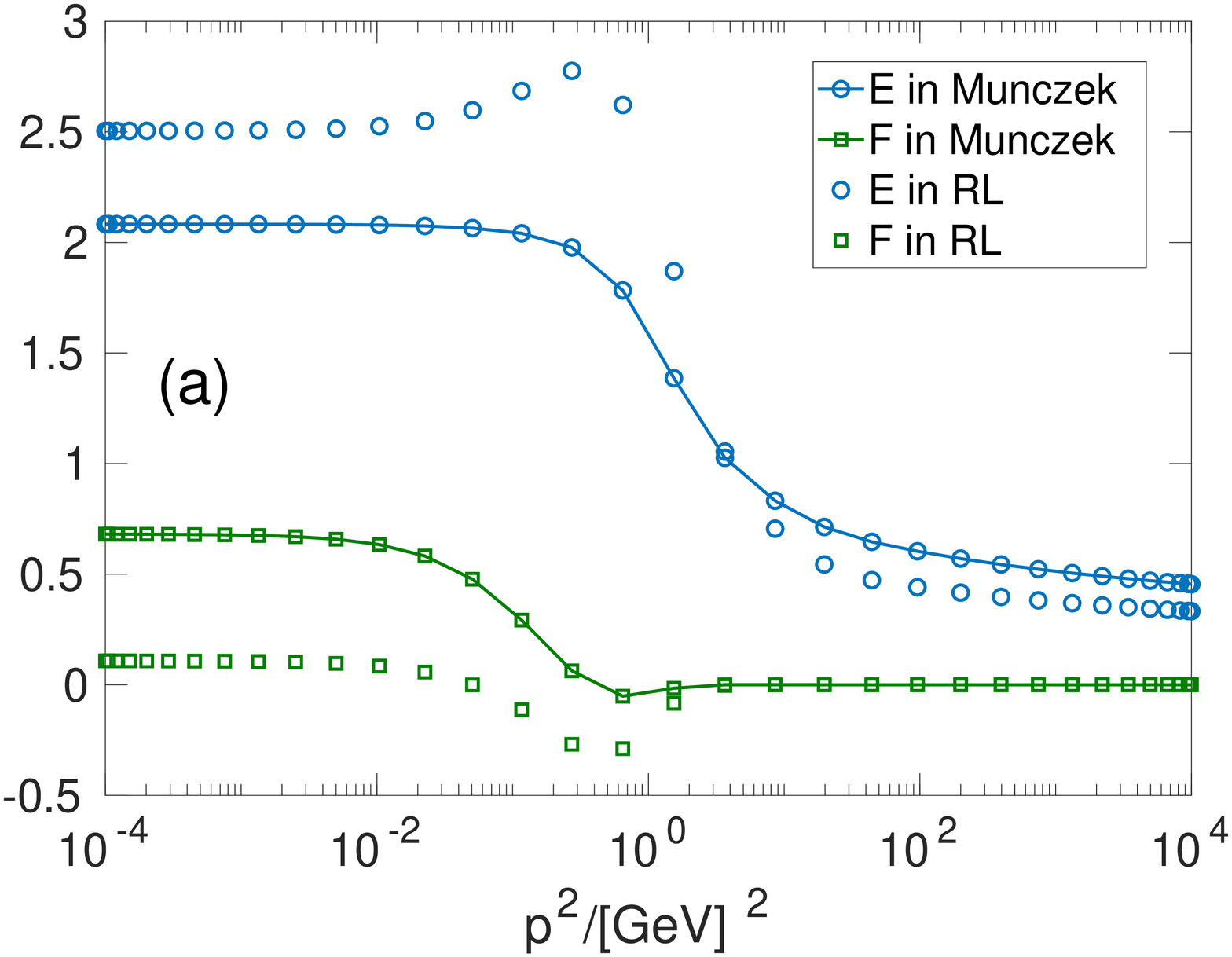}
%
%
\includegraphics[width=0.43\textwidth]{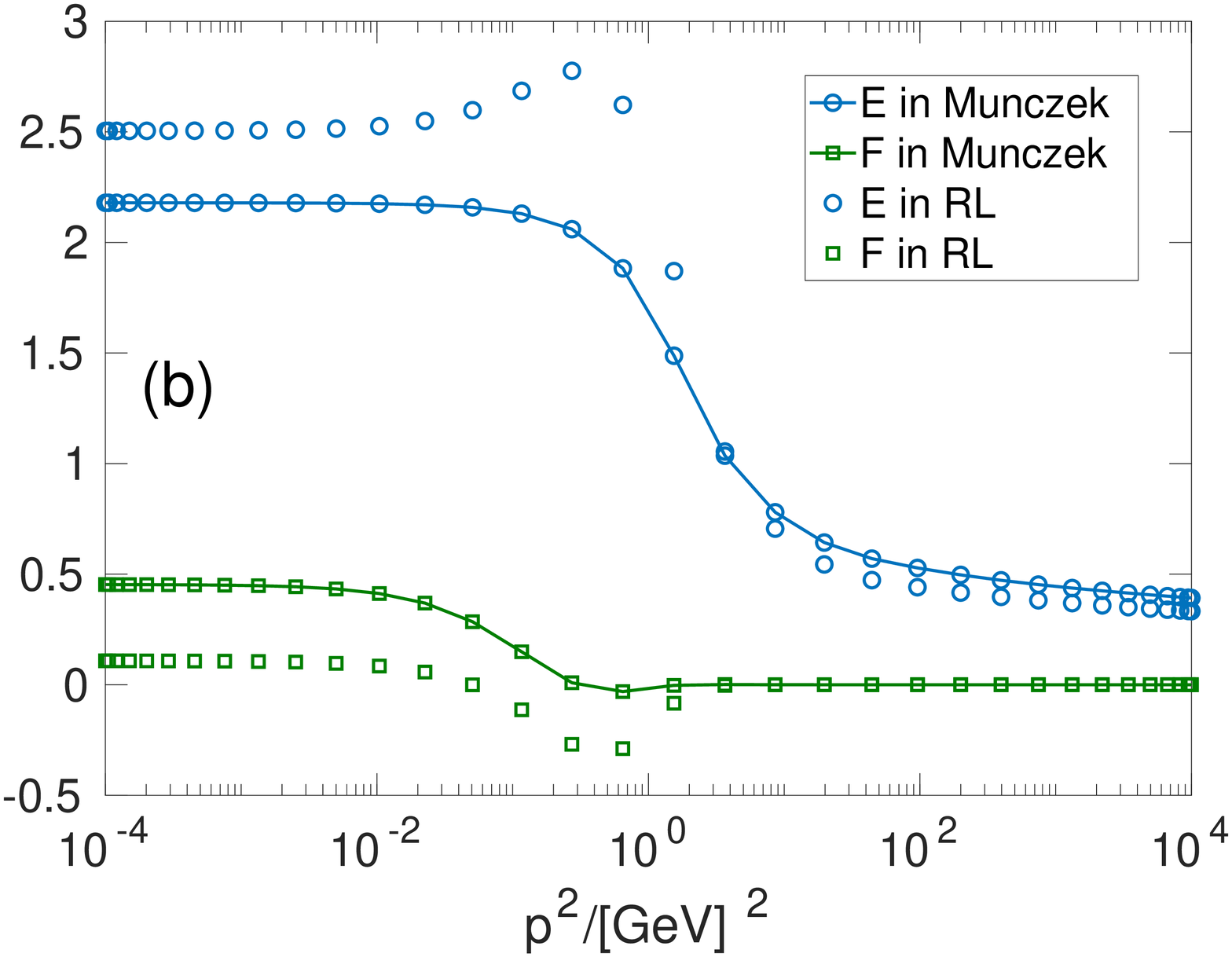}
\caption{Calculated quark scalar vertex solution under the RL approximation and
that under the Munczek's quark-gluon vertex model ((a)\ only the first scattering kernel in Munczek's approximation is taken into account.
(b)\ with the full scattering kernel in Munczek's approximation).
}
\label{fig-sv-sol}
\end{figure}

The corresponding full scalar vertex of the scalar distribution function is $\Gamma_{s}$,
whose tree level form is $1$. We decompose the full quark scalar vertex as
\begin{equation}
\begin{split}
\Gamma_{s} (k; 0) &= 1 * E(k^2) + i \slashed k * F(k^2)\,,
 \end{split}
 \label{eq-paraz-sv}
\end{equation}
in the Euclidean momentum space.
We calculate the dressed scalar vertex under the RL approximation and the Munczek's quark-gluon vertex model. The gluon parameters in our calculation are just those listed in Table~\ref{tab-gluon-para}.
The solutions of the dressed quark scalar vertex are showed in Fig.~\ref{fig-sv-sol}.
%
%
In consideration of the dressed effects of quark scalar vertex,
the result of dressed scalar distribution function in bag model should be rewritten from Eq.\eqref{s-dist-tree} as
\begin{equation}\label{s-dist-dressed}
\begin{split}
s_q^{}(x) &= \sqrt{2}M \left(\sum_\lambda \langle N; \bm 0; S| P_{f,\lambda}|N; \bm 0; S\rangle\right)\times\\
&\quad \int_{k_{min}}^\infty \frac{k dk}{(2\pi)^2} \bar\varphi(\bm k) \Gamma_s(k; 0) \varphi(\bm k) \frac{|\phi_2(\bm k)|^2}{|\phi_3(\bm 0)|^2}\,.
\end{split}
\end{equation}

As the same reason as that in calculating the dressed longitudinal distribution function,
the calculation of scalar distribution function in the bag model is independent of time.
We then derive the formula analytically in the Minkowskian space and take the dressed scalar functions from the DS equation calculation results which are calculated in the Euclidean space.

Next we give out the result for the  other term in Eq.~\eqref{eq-paraz-sv} as
\begin{equation}
\begin{split}
&\bar \varphi(\bm k) i \slashed k \varphi(\bm k) = \frac{4\pi R^3 \Omega^4}{\sqrt{2}(\Omega^2 - \sin^2 \Omega)} \times \\
&\qquad\left[k_0\left(t_{00}^2(k) + t_{11}^2(k)\right)-2k t_{00}(k) t_{11}(k)\right]\,.
\end{split}
\end{equation}
Noting that $k_{0} - k_{3} = M x,\;k_{0} = \Omega/R$, we get the dressed single quark scalar distribution function in a nucleon in the bag model as
\begin{equation}
\begin{split}
&s(x) = \frac{4\pi M R^3 \Omega^4}{(\Omega^2 - \sin^2 \Omega)}\left( \langle N; \bm 0| P_{f,m}|N; \bm 0\rangle\right)\times\\
&\left.\int_{k_{min}}^\infty \frac{k dk}{(2\pi)^2} \frac{|\phi_2(\bm k)|^2}{|\phi_3(\bm 0)|^2}\{\left[t_{00}^2(k) - t_{11}^2(k) \right] * E(k^2) \right.\\
&\left.+\left[k_0\left(t_{00}^2(k) + t_{11}^2(k)\right)-2k t_{00}(k) t_{11}(k)\right] * F(k^2)\right\}\,.
\end{split}
\end{equation}

We display the obtained dressed scalar distribution functions under the RL approximation and the Munczek's quark-gluon vertex model in Fig.~\ref{fig-scalar-dist-dress}.
The errors come from we vary the gluon parameter $\omega \in [0.49, 0.51]$ with which the pion decay constant is almost the same.
By integrating the scalar distribution function, we get the scalar charge in each of the cases
and obtained results are listed in Table~\ref{tab-scalarchar}.
In Table~\ref{tab-scalarchar} we list also the scalar charge result calculated from LQCD~\cite{Gupta:2018qil} for the convenience of comparison.
\begin{table}[h]
\caption{Calculated scalar charges in a proton under three different cases: the bare scalar vertex, the dressed scalar vertex in RL approximation and the dressed scalar vertex in Munczek's quark-gluon vertex model. We also list the scalar charge from the LQCD results~\cite{Gupta:2018qil}.}
\label{tab-scalarchar}
\begin{ruledtabular}
\begin{tabular}{c | c}
~~Approximation~~ & ~~~~~~ Scalar Charge $g_{s}^{}$~~~~~~ \\
\hline
bare & 0.6655\\
\hline
RL & $1.794^{+(39)}_{-(40)}$\\
\hline
Mun. $1_{\text{st}}$ & $1.406^{+(26)}_{-(5)}$\\
\hline
Mun. full & $1.443^{+(42)}_{-(42)}$\\
\hline
LQCD & 1.022(80)(60) \\
\end{tabular}
\end{ruledtabular}
\end{table}

We can see from Fig.~\ref{fig-scalar-dist-dress} and Table~\ref{tab-scalarchar} that, when considering the dressed effects on the scalar distribution function, the distribution gets much higher than that in bare scalar vertex approximation.
Furthermore, comparing the results with different dressed effects, one can notice that,
the scalar charge in the case beyond RL approximation (Munczek's quark-gluon vertex model) is much smaller  than that in the RL approximation. The scalar distribution and the scalar charge are quite close, respectively, in the cases of Mun. $1_{\text{st}}$ and the Mun. full truncation.
We may then infer that, when we take more and more dressed effects into account,
the scalar charges approach gradually to a value.

\section{Summary and Remarks}
\label{sec-sum}
In this work, we gave out a bridge that connects the  nonperturbative QCD calculation in Euclidean space
(especially, via the DS equation approach) and the parton distribution functions in Minkowskian space.
We combine the MIT bag model with the DS equation approach to calculate the dressed distribution functions
in a nucleon, such as the longitudinal distribution, the transversity distribution and the scalar distribution.
We obtain the dressed axial charge, tensor charge as well as the scalar charge from the distribution functions.

We compare the obtained distribution functions with or without the dressed effects.
Furthermore, we compare the dressed effects on the distribution functions under different truncation schemes of the DS equations, the RL approximation and the Munczek's quark-gluon vertex model.
We add more dressed effects beyond the RL approximation step by step to explore their influences on the dressed quark distribution functions.

For the longitudinal distribution function $g(x)$, we find that the axial coupling $g_{A}^{}$ calculated in bare axial vector vertex is close to the experimental value.
The dressed effects lower the axial coupling a bit in the RL approximation comparing to that in the bare axial vector vertex.
By comparing the dressed effect in the RL approximation and that in the cases of beyond the RL approximation (Munczek's quark-gluon vertex approximation), we see that the longitudinal distribution function rises up gradually as more dressed effects are taken into account.

For the transversity distribution function $h(x)$, we find that the dressed effects are significant.
The dressed effects lower the transversity distribution and decrease the tensor charges in the RL approximation comparing to that, respectively, in the bare tensor vertex.
The transversity distribution function gets decreased continuously in the Mun. $1_{\text{st}}$ truncation (considering only the first scattering kernel) contrast to that in the RL approximation and rises up in the Mun. full truncation (restoring the full two scattering kernels) comparing to that in the Mun. $1_{\text{st}}$ truncation.

For the scalar distribution function $s(x)$, we find that the dressed effects are also important.
The dressed effects rises the scalar distribution much and so do to the scalar charges.
By comparing the dressed effects in the RL approximation and in the cases beyond the RL approximation (Munczek's quark-gluon vertex approximation), we find that the scalar distribution function gets lowered
in the Mun. $1_{\text{st}}$ and in the Mun. full truncation than that in the RL approximation.
However the scalar distributions in the Mun. $1_{\text{st}}$ truncation and in the Mun. full truncation
are quite close to each other.
In this sense, we may deduce that, when more and more dressed effects are taken into account,
the scalar coupling decreases gradually and approaches finally to a definite value.

From these three examples, we can observe that combining the MIT bag model with the DS equation approach
in the Euclidean space is a quite simple and efficient way to look insight into the dressed effects on the distribution functions
in a nucleon in Minkowskian space qualitatively.
Actually, we do not need to limit to the DS equation approach in the nonperturbative QCD calculations.
All the approaches that compute the dressed effects in Euclidean space can be combined in a similar way.
This may be a convenient and efficient way to connect the two aspects.

\begin{acknowledgments}

We are grateful to the selfless help of professor Wouter. J. Waalewijn. Work supported by:
the National Natural Science Foundation of China under contracts No. 11435001, and No. 11775041, the National Key Basic Research Program of China under contract No. 2015CB856900, and
LC thanks also the Chinese Government's \emph{Thousand Talents Plan for Young Professionals}.

\end{acknowledgments}


%

\appendix

\section{Bag Wave Function in Momentum Space}
\label{appendix-wave-fourier}
The distribution fucntions are defined in the momentum space, so we need the wave function in the momentum space:
\begin{equation}
\varphi(\bm k, t) = \int d^3x e^{i \bm k \cdot \bm x} \varphi(\bm x, t)
\end{equation}
The result is
\begin{equation}
\varphi(\bm k, t) = 4\pi R^3 N
\begin{bmatrix}
t_{00}(y) U_m\\
\bm \sigma \cdot \hat{\bm k} t_{11}(y) U_m
\end{bmatrix}
e^{-i \frac{\Omega t}{R}}\,,
\end{equation}
where $y=k R, k = |\bm k|, \hat{\bm k} = \bm k/k$ and
\begin{equation}
t_{i\, j}(y) = \int_0^1 du u^2 j_i(yu) j_j(\Omega u)\,, \qquad y = kR\,.
\end{equation}
Specifically,
\begin{subequations}
\begin{align}
&t_{00}(y) = \int_0^1 du u^2 j_0(u y) j_0(u \Omega) \\
\nonumber&\qquad\;\;= \frac{1}{2 \omega y}\left[\frac{\sin(y-\Omega)}{y-\Omega}-\frac{\sin(y+\Omega)}{y+\Omega}\right]\,,\\
&t_{11}(y) = \int_0^1 du u^2 j_1(u y) j_1(u \Omega) \\
\nonumber&= \frac{-2\sin(y)\sin(\Omega)}{2\Omega^2 y^2}+\frac{1}{2 \Omega y}\left[\frac{\sin(y+\Omega)}{y+\Omega}+\frac{\sin(y-\Omega)}{y-\Omega}\right]\,.
\end{align}
\end{subequations}

\section{Bag Wave Function Overlap Integral}
\label{appendix-overlap-int}
The calculation of bag wave function overlap integral is tricky and  we show the results here.
We denote the bag wave function  overlap integral as
\begin{equation}
I = \int d^3y \varphi^\dagger(\bm y - \bm x) \varphi(\bm y)\,.
\end{equation}
The result is
\begin{eqnarray}
\begin{split}
I &= \frac{\Omega}{2 v (\Omega^2 - \sin^{2} \Omega)}
\left[ \Big{(} \Omega - \frac{1-\cos{2}\Omega}{2\Omega} - v \Big{)} \sin 2v \right. \quad \\
&\qquad \left. - \Big{(} \frac{1}{2} + \frac{\sin 2\Omega}{2\Omega} \Big{)} \cos 2v
+ \frac{1}{2} + \frac{\sin 2\Omega}{2\Omega} \right. \\
& \qquad \left. - \frac{1 - \cos 2\Omega}{2\Omega^{2}} v^{2} \right] \\
& \xlongequal{\triangle} \frac{\Omega}{2 v (\Omega^2 - \sin^2 \Omega)} T(v)\, ,
\end{split}
\end{eqnarray}
where $v = \frac{\Omega x}{2 R},\, x = |\bm x|$.
Let $|\bm p| R = u$, then
\begin{eqnarray}
\begin{split}
&|\phi_n(\bm p)|^2 \\
&= \int d^3a  e^{-i \bm p \cdot \bm a} \left[ \int d^3x \varphi^\dagger (\bm x - \bm a) \varphi(\bm x) \right]^n\,,\\
&= 2\pi \int_0^R da a^2 \int_{-1}^1 d(-\cos \theta) e^{- i \frac{2 u v}{\Omega} \cos \theta} \\
& \quad \times \Big{[} \frac{\Omega}{2 v (\Omega^2 - \sin^2 \Omega)} \Big{]}^{n} T^{n}(v) \\
&= \frac{2^{4-n} \pi R^3 \Omega^{n-2}}{u \left(\Omega^2 - \sin^2 \Omega\right)^n}\int_0^{\Omega/2} \frac{d v}{v^{n-1}} \sin \frac{2 u v}{\Omega} T^n(v)\,.
\end{split}
\end{eqnarray}
Specifically,
\begin{subequations}
\begin{align}
&|\phi_2(\bm p)|^2 = \frac{4\pi R^3}{u(\Omega^2 - \sin^2 \Omega)^2}\int_0^{\Omega/2} \frac{d v}{v} \sin \frac{2 u v}{\Omega} T^{2}(v) \, , \\
& \nonumber \\
&|\phi_3(\bm 0)|^2 = \frac{4\pi R^3}{(\Omega^2 - \sin^2 \Omega)^3}\int_0^{\Omega/2} \frac{d v}{v}  T^3(v)\,.
\end{align}
\end{subequations}

\end{document}